\documentstyle[12pt,epsfig]{article}
\def\setboxz@h{\setbox\z@\hbox}
\def\wdz@{\wd\z@}

\expandafter\chardef\csname pre amssym.def at\endcsname=\the\catcode`\@
\catcode`\@=11
\def\undefine#1{\let#1\undefined}
\def\newsymbol#1#2#3#4#5{\let\next@\relax
 \ifnum#2=\@ne\let\next@\msafam@\else
 \ifnum#2=\tw@\let\next@\msbfam@\fi\fi
 \mathchardef#1="#3\next@#4#5}
\def\mathhexbox@#1#2#3{\relax
 \ifmmode\mathpalette{}{\m@th\mathchar"#1#2#3}%
 \else\leavevmode\hbox{$\m@th\mathchar"#1#2#3$}\fi}
\def\hexnumber@#1{\ifcase#1 0\or 1\or 2\or 3\or 4\or 5\or 6\or 7\or 8\or
 9\or A\or B\or C\or D\or E\or F\fi}
\font\tenmsa=msam10
\font\sevenmsa=msam7
\font\fivemsa=msam5
\newfam\msafam
\textfont\msafam=\tenmsa
\scriptfont\msafam=\sevenmsa
\scriptscriptfont\msafam=\fivemsa
\edef\msafam@{\hexnumber@\msafam}
\mathchardef\dabar@"0\msafam@39
\def\dashrightarrow{\mathrel{\dabar@\dabar@\mathchar"0\msafam@4B}}
\def\dashleftarrow{\mathrel{\mathchar"0\msafam@4C\dabar@\dabar@}}

\def\ulcorner{\delimiter"4\msafam@70\msafam@70 }
\def\urcorner{\delimiter"5\msafam@71\msafam@71 }
\def\llcorner{\delimiter"4\msafam@78\msafam@78 }
\def\lrcorner{\delimiter"5\msafam@79\msafam@79 }
\def\yen{{\mathhexbox@\msafam@55 }}
\def\checkmark{{\mathhexbox@\msafam@58 }}
\def\circledR{{\mathhexbox@\msafam@72 }}
\def\maltese{{\mathhexbox@\msafam@7A }}
\font\tenmsb=msbm10
\font\sevenmsb=msbm7
\font\fivemsb=msbm5
\newfam\msbfam
\textfont\msbfam=\tenmsb
\scriptfont\msbfam=\sevenmsb
\scriptscriptfont\msbfam=\fivemsb
\edef\msbfam@{\hexnumber@\msbfam}

\def\widehat#1{\setbox\z@\hbox{$\m@th#1$}%
 \ifdim\wd\z@>\tw@ em\mathaccent"0\msbfam@5B{#1}%
 \else\mathaccent"0362{#1}\fi}
\def\widetilde#1{\setbox\z@\hbox{$\m@th#1$}%
 \ifdim\wd\z@>\tw@ em\mathaccent"0\msbfam@5D{#1}%
 \else\mathaccent"0365{#1}\fi}
\font\teneufm=eufm10
\font\seveneufm=eufm7
\font\fiveeufm=eufm5
\newfam\eufmfam
\textfont\eufmfam=\teneufm
\scriptfont\eufmfam=\seveneufm
\scriptscriptfont\eufmfam=\fiveeufm

\catcode`\@=\csname pre amssym.def at\endcsname
\expandafter\ifx\csname pre amssym.tex at\endcsname\relax
                                         \else \endinput\fi
\expandafter\chardef\csname pre amssym.tex at\endcsname=\the\catcode`\@
\catcode`\@=11
\newsymbol\gtrsim 1326
\newsymbol\lesssim 132E
\catcode`\@=\csname pre amssym.tex at\endcsname
\newcommand{\Rp}{\mbox{$\not \hspace{-0.15cm} R_p$}}
\newcommand{\cm}{\mbox{\rm ~cm}}
\def\GeV{\hbox{$\;\hbox{\rm GeV}$}}
\newcommand{\picob}{\mbox{{\rm ~pb}~}}

\begin{document}
\begin{titlepage}
%
\vspace*{4.cm}
\begin{center}
\begin{Large}
\boldmath
\bf{A Search for Squarks of Rp-Violating SUSY at HERA\\}
\unboldmath
 
\vspace*{2.cm}
H1 Collaboration \\
\end{Large}
 
\vspace*{1cm}
 
\end{center}
 
\vspace*{1cm}
 
\begin{abstract}
\noindent
A search for squarks of $R$-parity violating supersymmetry is
performed in $ep$ collisions at HERA using H1 1994 $e^+$ data.
Direct single production of squarks of each generation by  $e^+$-quark
fusion via a Yukawa coupling $\lambda'$ is considered.
All possible $R$-parity violating decays and gauge decays of the
squarks are taken into account.
No significant deviation from the Standard Model predictions
is found in the various multi-lepton and multi-jet final
states studied and exclusion limits are derived.
At $95\%$ confidence level, the existence of first generation
squarks is excluded for masses up to $240 \GeV $ for coupling values
$\lambda' \gtrsim \sqrt{ 4\pi \alpha_{em}}$.
The limits obtained are shown to be only weakly dependent on the
free parameters of the Minimal Supersymmetric Standard Model.
Stop squarks are excluded for masses up to $138 \GeV$
for coupling $\lambda' \times \cos \theta_t$ to $e^+d$ pairs
$\gtrsim 0.1 \times \sqrt{ 4\pi \alpha_{em}}$,
where $\theta_t$ is the mass mixing angle.
 
Light stop squarks are furthermore searched for through pair production
in $\gamma$-gluon fusion processes.
No signal is observed and exclusion limits are derived.
Masses in the range $9$ to $24.4 \GeV$ are excluded at $95\%$
confidence level for $\lambda' \times \cos \theta_t > 10^{-4}$.
\vspace{1cm}
 
 
\end{abstract}
\end{titlepage}
 
\vfill
\clearpage
 
%
\begin{flushleft}
 S.~Aid$^{14}$,                   
 V.~Andreev$^{26}$,               
 B.~Andrieu$^{29}$,               
 R.-D.~Appuhn$^{12}$,             
 M.~Arpagaus$^{37}$,              
 A.~Babaev$^{25}$,                
 J.~B\"ahr$^{36}$,                
 J.~B\'an$^{18}$,                 
 Y.~Ban$^{28}$,                   
 P.~Baranov$^{26}$
% ...
\end{flushleft}
\begin{flushleft} {\it
% ...
$ ^{12}$ DESY, Hamburg, Germany$^a$ \\
% ...
$ ^{14}$ II. Institut f\"ur Experimentalphysik, Universit\"at Hamburg,
                                     Hamburg, Germany$^ a$  \\
% ...
$ ^{18}$ Institute of Experimental Physics, Slovak Academy of
                Sciences, Ko\v{s}ice, Slovak Republic$^ f$\\
% ...
$ ^{25}$ Institute for Theoretical and Experimental Physics,
                                                 Moscow, Russia \\
 $ ^{26}$ Lebedev Physical Institute, Moscow, Russia$^ f$ \\
% ...
$ ^{28}$ LAL, Universit\'{e} de Paris-Sud, IN2P3-CNRS,
                            Orsay, France\\
 $ ^{29}$ LPNHE, Ecole Polytechnique, IN2P3-CNRS,
                             Palaiseau, France \\
% ...
$ ^{36}$ DESY, Institut f\"ur Hochenergiephysik,
                              Zeuthen, Germany$^ a$\\
 $ ^{37}$ Institut f\"ur Teilchenphysik,
          ETH, Z\"urich, Switzerland$^ i$\\
% ...
$ ^a$ Supported by the Bundesministerium f\"ur
        Forschung und Technologie, FRG,
        under contract numbers 6AC17P, 6AC47P, 6DO57I, 6HH17P, 6HH27I,
        6HD17I, 6HD27I, 6KI17P, 6MP17I, and 6WT87P \\
% ...
$ ^f$ Supported by the Deutsche Forschungsgemeinschaft\\
% ...
$ ^i$ Supported by the Swiss National Science Foundation\\
 }
\end{flushleft}
 
\newpage
\section{Introduction}
\label{sec:intro}
 
The search for squarks, the scalar supersymmetric (SUSY) partners of
the quarks, is especially promising at the $ep$ collider HERA if they
possess a lepton number violating Yukawa coupling $\lambda'$ to
lepton--quark pairs.
Such squarks, present in the $R$-parity violating (\Rp) SUSY extension
of the Standard Model (SM), can be singly produced via the coupling
$\lambda'$ as $s$-channel resonances.
Masses up to the kinematic limit of $\sqrt{s} \simeq 300 \GeV$ are
accessible by the fusion of the $27.5 \GeV$ initial state positron
with a quark of the $820 \GeV$ incoming proton.
In the low mass range, pair production via $\gamma$-gluon fusion provides
a complementary search largely insensitive to the Yukawa coupling.
 
In this paper, squarks are searched through single production via a
\Rp\ coupling, considering both \Rp\ decays and all possible decays via
gauge couplings involving mixed states of gauginos and higgsinos.
A search for pair production of light stops at low masses via
$\gamma$-gluon fusion is also carried out.
The analysis uses the 1994 $e^+ p$ data corresponding to an
integrated luminosity of ${\cal L}_{data} = 2.83 \picob^{-1}$.
Earlier squark searches at HERA were presented in~\cite{H1LQ94}.
 
\section{Phenomenology}
\label{sec:pheno}
 
The general SUSY superpotential allows for gauge invariant terms with
Yukawa couplings between the scalar squarks ($\tilde{q}$) or sleptons
($\tilde{l}$) and the known SM fermions.
Such couplings exist if one assumes the possibility of violating
(multiplicatively) the conservation of $R$-parity which is imposed
in the Minimal Supersymmetric Standard Model (MSSM);
$R_p\,=\,(-1)^{3B+L+2S}$
where $S$ denotes the spin, $B$ the baryon number and $L$ the lepton
number of the particles.
Of particular interest for HERA are the \Rp\ terms
$\lambda'_{ijk} L_{i}Q_{j}\bar{D}_k$ of the superpotential
which allow for lepton number violating processes.
By convention the $ijk$ indices correspond to the generations of
the superfields $L_{i}$, $Q_{j}$ and $\bar{D}_k$
containing respectively the left-handed lepton doublet, quark doublet
and the right handed quark singlet.
Expanded in terms of matter fields, the interaction Lagrangian
reads~\cite{RPVIOLATION} :
\begin{eqnarray}
{\cal{L}}_{L_{i}Q_{j}\bar{D_{k}}} &=
   & \lambda^{\prime}_{ijk}
              \left[ -\tilde{e}_{L}^{i} u^j_L \bar{d}_R^k
              - e^i_L \tilde{u}^j_L \bar{d}^k_R - (\bar{e}_L^i)^c u^j_L
     \tilde{d}^{k*}_R \right.           \nonumber \\
 \mbox{} &\mbox{}
 & \left. + \tilde{\nu}^i_L d^j_L \bar{d}^k_R + \nu_L \tilde{d}^j_L
    \bar{d}^k_R + (\bar{\nu}^i_L)^c d^j_L \tilde{d}^{k*}_R \right]
   +\mbox{h.c.}             \nonumber
 \nonumber
\end{eqnarray}
where the superscripts $^c$ denote the charge conjugate spinors
and the $^*$ the complex conjugate of scalar fields.
For the scalars the `R' and `L' indices distinguish independent
fields describing superpartners of right- and left-handed fermions.
Hence, with an $e^+$ in the initial state, the couplings
$\lambda'_{1jk}$ allow for resonant production of squarks through
positron-quark fusion.
The list of possible single production processes is given in
table~\ref{tab:sqprod}.
%
%
\begin{table*}[htb]
  \renewcommand{\doublerulesep}{0.4pt}
  \renewcommand{\arraystretch}{1.2}
 \begin{center}
 \begin{tabular}{p{0.40\textwidth}p{0.60\textwidth}}
         \caption
         {\small \label{tab:sqprod}
         Squark production processes at HERA ($e^+$ beam)
         via a $R$-parity violating
         $\lambda'_{1jk}$ coupling.} &
   \begin{tabular}{||c||c|c||}
   \hline \hline
   $\lambda'_{1jk}$ & \multicolumn{2}{c||}{production process} \\
   \hline
   111 & $e^+ +\bar{u} \rightarrow \bar{\tilde{d}_R}$
       &$e^+ +d \rightarrow \tilde{u}_L $\\
   112 & $e^+ +\bar{u} \rightarrow \bar{\tilde{s}_R}$
       &$e^+ +s \rightarrow \tilde{u}_L $\\
   113 & $e^+ +\bar{u} \rightarrow \bar{\tilde{b}_R}$
       &$e^+ +b \rightarrow \tilde{u}_L $\\
   121 & $e^+ +\bar{c} \rightarrow \bar{\tilde{d}_R}$
       &$e^+ +d \rightarrow \tilde{c}_L $\\
   122 & $e^+ +\bar{c} \rightarrow \bar{\tilde{s}_R}$
       &$e^+ +s \rightarrow \tilde{c}_L $\\
   123 & $e^+ +\bar{c} \rightarrow \bar{\tilde{b}_R}$
       &$e^+ +b \rightarrow \tilde{c}_L $\\
   131 & $e^+ +\bar{t} \rightarrow \bar{\tilde{d}_R}$
       &$e^+ +d \rightarrow \tilde{t}_L $\\
   132 & $e^+ +\bar{t} \rightarrow \bar{\tilde{s}_R}$
       &$e^+ +s \rightarrow \tilde{t}_L $\\
   133 & $e^+ +\bar{t} \rightarrow \bar{\tilde{b}_R}$
       &$e^+ +b \rightarrow \tilde{t}_L $\\
   \hline \hline
  \end{tabular}
  \end{tabular}
\end{center}
\end{table*}
In this paper, the squark search is carried with the simplifying
assumptions that:
\begin{itemize}
\item only one of the $\lambda'_{1jk}$ dominates;
\item squarks ($\tilde{q}_R$ and $\tilde{q}_L$) of the first and 
      second generation are quasi-degenerate in mass (the case 
      of the stop squark is considered separately);
\item the lightest supersymmetric particle is the lightest
      neutralino $\chi_1^0$;
\item gluinos are heavier than the squarks such that decays
      $\tilde{q} \rightarrow q + \tilde{g}$
      are kinematically forbidden.
\end{itemize}
The squarks decay either via their Yukawa coupling into fermions,
or via their gauge couplings into a quark and either a neutralino
$\chi_i^0$ ($i=1,4$) or a chargino $\chi_j^{+}$ ($j=1,2$).
The mass eigenstates $\chi_i^0$ and $\chi_j^{+}$ are mixed
states of gauginos and higgsinos and are in general unstable.
In contrast to the MSSM, this also holds in \Rp\ SUSY for the lightest
supersymmetric particle (LSP) which decays via $\lambda'_{1jk}$
into a quark, an antiquark and a lepton~\cite{RPVIOLATION}.
 
\begin{figure}[htb]
\vspace{-1.5cm}
 
  \begin{center}
     \mbox{\epsfxsize=0.9\textwidth \epsffile{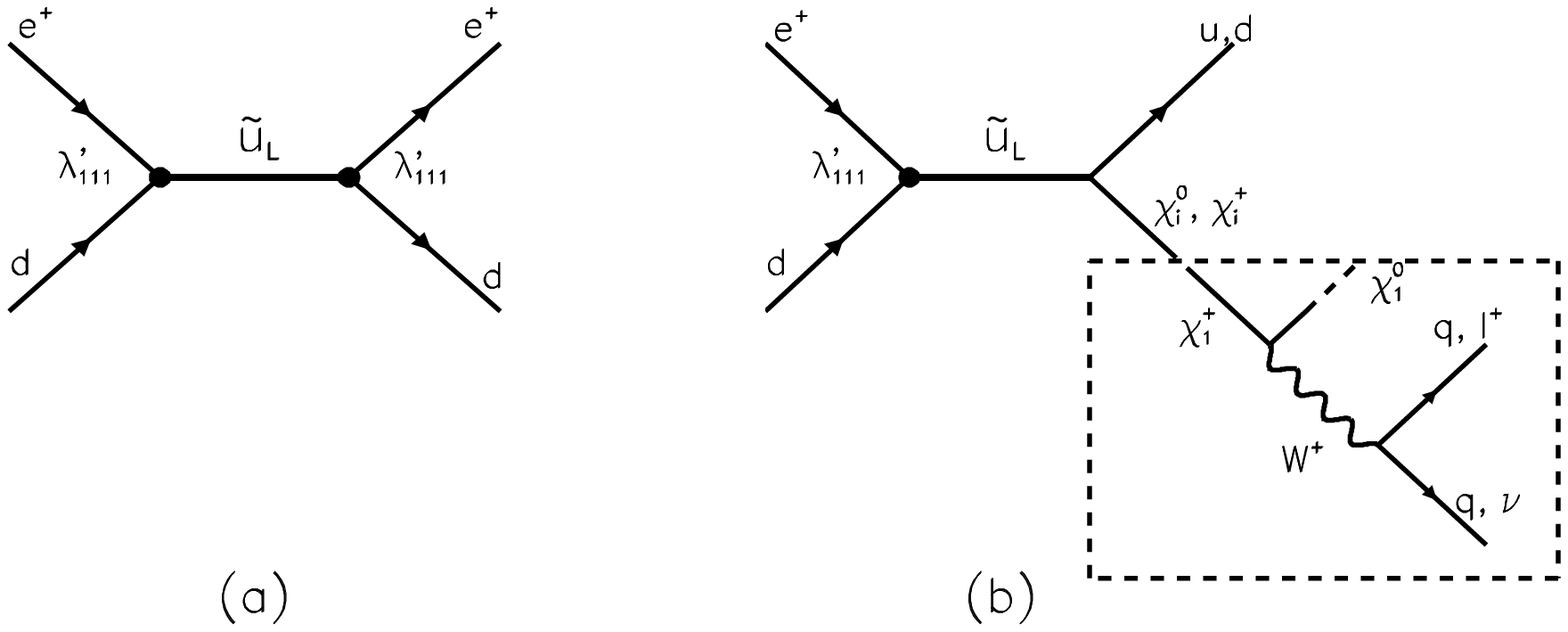}}
 
     \vspace{-2.1cm}
 
     \mbox{\epsfxsize=0.9\textwidth \epsffile{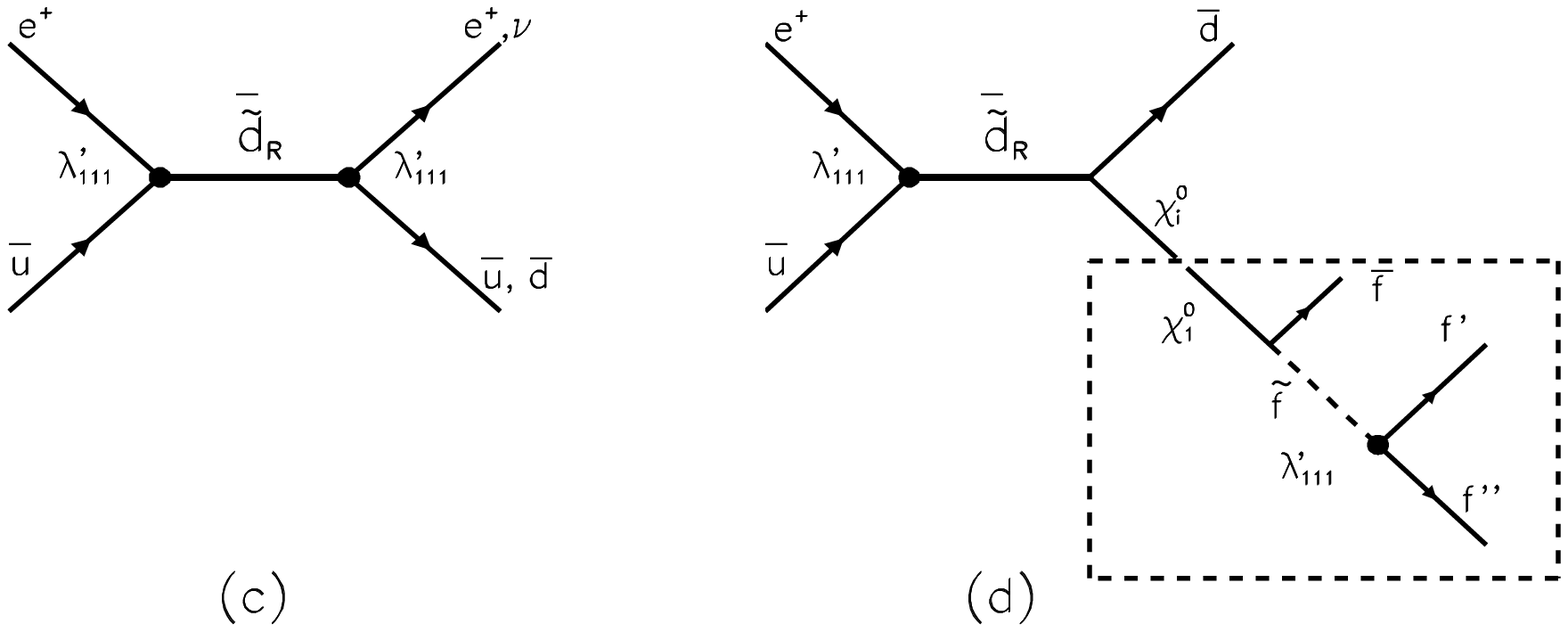}}
  \end{center}
 
\vspace{-1.5cm}

 \caption[]{ \label{fig:sqdiag}
    {\small Lowest order $s$-channel diagrams for first generation
      squark production at HERA followed by
      (a),(c) \Rp\ decays and (b),(d) gauge decays.
      In (b) and (d), the emerging neutralino or chargino might
      subsequently undergo \Rp\ decays of which examples are
      shown in the doted boxes for (b) the $\chi_1^{+}$ and
      (d) the $\chi_1^0$. }}
\end{figure}

Typical diagrams for the production of first generation squarks are
shown in Fig.~\ref{fig:sqdiag}.
%
By gauge symmetry only the $\bar{\tilde{d}}_R$ and $\tilde{u}_L$ are
produced via the $\lambda'$ couplings.
These 
have in general widely different allowed or dominant decay modes.
 
 
In cases where both production and decay occur through a
$\lambda'_{1jk}$ coupling (e.g. Fig.~\ref{fig:sqdiag}a and c for
$\lambda'_{111} \ne 0$), the squarks behave as scalar
leptoquarks~\cite{H1LQ95,BUCHMULL}.
For $\lambda'_{111} \ne 0$, the $\bar{\tilde{d}}_R$ resemble the
$\bar{S^0}$ leptoquark and decays in either $e^+ + \bar{u}$ or
$\nu_e + \bar{d}$  while the $\tilde{u}_L$ resemble the
$\bar{\tilde{S}}_{1/2}$ and only decays into $e^+ \bar{d}$.
Hence, the final state signatures consist of a lepton and a jet and
are, event-by-event, indistinguishable from the SM neutral
(NC) and charged current (CC) deep inelastic scattering (DIS).
The strategy is then to look for resonances in DIS--like events
at high mass, exploiting the characteristic angular distribution
of the decay products expected for a scalar particle.
 
 
In cases where the squark decay occurs through gauge couplings
(e.g. Fig.~\ref{fig:sqdiag}b and d), one has to consider for the
$\tilde{u}_L$ the processes $\tilde{u}_L \rightarrow u \chi_i^0$ or
$d \chi_j^+$ while for the $\bar{\tilde{d}}_R$
only $\bar{\tilde{d}}_R \rightarrow \bar{d} \chi_i^0$ is allowed.
This is because the $SU(2)_L$ symmetry which implies in the SM
that the right handed fermions do not couple to the $W$ boson
also forbids a coupling of $\bar{\tilde{d}}_R$ to the $\tilde{W}$.
Hence, the $\bar{\tilde{d}}_R$ can only weakly couple (in proportion
to the $d$ quark mass) to the $\chi_j^+$ through its higgsino component.
 
%
%
\begin{table}[htb]
 \renewcommand{\doublerulesep}{0.4pt}
 \renewcommand{\arraystretch}{1.0}
 \begin{center}
  \begin{tabular}{||c|c|l|l||}
  \hline \hline
  Channel  & $\chi_1^0$ & \multicolumn{1}{c|}{Decay processes}
                        & \multicolumn{1}{c||}{Signature} \\
           & nature & \multicolumn{1}{c|}{ }
                    & \multicolumn{1}{c||}{ }         \\ \hline
  S1 &  $\tilde{\gamma}$,$\tilde{Z}$,$\tilde{H}$
     &  \begin{tabular}{cccccc}
          $\tilde{q}$ & $\stackrel{\lambda'}{\longrightarrow}$
                      & $e^+$   & $q'$    &    &
        \end{tabular}
     &  \begin{tabular}{l}
        High $P_T$ $e^+$ + 1 jet
        \end{tabular} \\                                        \hline
  S2 &  \begin{tabular}{c}
          $\tilde{\gamma}$,$\tilde{Z}$,$\tilde{H}$ \\
          $\tilde{H}$
        \end{tabular}
     &  \begin{tabular}{cccccc}
          $\bar{\tilde{d}}_R$ & $\stackrel{\lambda'}{\longrightarrow}$
                      & $\nu_e$   & $\bar{d}$     &    & \\
          $\tilde{q}$         & $\longrightarrow$
                      & $q$       & $\chi_1^0$    &    &
        \end{tabular}
     &  \begin{tabular}{l}
         Missing $P_T$ + 1 jet
        \end{tabular} \\                                        \hline
  S3 & \begin{tabular}{c}
           $\tilde{\gamma}$,$\tilde{Z}$ \\ \\
           $\tilde{\gamma}$,$\tilde{Z}$,$\tilde{H}$ \\ \\
           $\tilde{\gamma}$,$\tilde{Z}$ \\ \\ \\ \\ \vspace{-0.5cm} \\
       \end{tabular}
     & \begin{tabular}{ccccll}
         $\tilde{q}$ & $\longrightarrow$
                     & $q$ & $\chi_1^0$  &  &  \\
         &  &        & $\stackrel{\lambda'}{\hookrightarrow}$
                     & $e^+ \bar{q}' q''$ & \\
         $\tilde{u}_L$ & $\longrightarrow$ & $d$ & $\chi_1^+$ &  & \\
         &  &        & $\stackrel{\lambda'}{\hookrightarrow}$
                     & $e^+ d \bar{d}$ &  \\
         $\tilde{u}_L$ & $\longrightarrow$
                     & $d$ & $\chi_1^+$ &  & \\
         &  &        & $\hookrightarrow$
                     & $W^+$ & $\hspace{-1.1cm}\chi_1^0$ \\
         &  &  &  & $\:|$
                  & $\hspace{-1.1cm}
                \stackrel{\lambda'}{\hookrightarrow}$ $e^+ \bar{q}' q''$
                \vspace{-0.2cm} \\
         &  &  &  & $\:\mid$ \vspace{-0.3cm} &  \\
         &  &  &  & $\:\rightarrow$  $q \:\: \bar{q}'$  &
       \end{tabular}
     & \begin{tabular}{l}
        High $P_T$ $e^+$ \\
        + multiple jets
       \end{tabular}\\                                          \hline
  S4 & \begin{tabular}{c}
          $\tilde{\gamma}$,$\tilde{Z}$ \\ \\
          $\tilde{\gamma}$,$\tilde{Z}$ \\ \vspace{-0.5cm} \\ \\ \\  \\
       \end{tabular}
     & \begin{tabular}{ccccll}
         $\tilde{q}$ & $\longrightarrow$
                     & $q$ & $\chi_1^0$ &  & \\
         &  &        & $\stackrel{\lambda'}{\hookrightarrow}$
         & $e^- \bar{q}' q''$ &  \\
         $\tilde{u}_L$ & $\longrightarrow$ & $d$ & $\chi_1^+$ &  & \\
         &  &        & $\hookrightarrow$
                     & $W^+$ & $\hspace{-1.1cm}\chi_1^0$ \\
         &  &  & & $\:|$
         & $\hspace{-1.1cm}
           \stackrel{\lambda'}{\hookrightarrow}$ $e^- \bar{q}' q''$
           \vspace{-0.2cm} \\
         &  &  &     & $\:\mid$ \vspace{-0.3cm}  &  \\
         &  &  &     & $\:\rightarrow$  $q \:\: \bar{q}'$  &
       \end{tabular}
     & \begin{tabular}{l}
        High $P_T$ $e^-$ \\ (i.e. wrong sign lepton) \\
        + multiple jets
       \end{tabular}\\
   \hline \hline
  \end{tabular}
  \caption[]
          {\small \label{tab:sqtopo1}
               Squark decay channels in \Rp\ SUSY classified per
               distinguishable event topologies (first part).
               The dominant component of the $\chi_1^0$ for which a
               given decay chain is relevant is given in the second
               column.
               The list of processes contributing to a given event
               topology is here representative but not exhaustive,
               e.g. the gauge decays of the $\chi_1^+$ involving a
               virtual $W^+$ (Fig.~\ref{fig:sqdiag}b)
               may also proceed via a virtual sfermion.}
 \end{center}
\end{table}
\begin{table}[htb]
 \renewcommand{\doublerulesep}{0.4pt}
 \renewcommand{\arraystretch}{1.0}
 \begin{center}
  \begin{tabular}{||c|c|l|l||}
  \hline \hline
  Channel  & $\chi_1^0$ & \multicolumn{1}{c|}{Decay processes}
                        & \multicolumn{1}{c||}{Signature} \\
           & nature & \multicolumn{1}{c|}{ }
                    & \multicolumn{1}{c||}{ }         \\ \hline
%
  S5 & \begin{tabular}{c}
          $\tilde{\gamma}$,$\tilde{Z}$ \\ \\
          $\tilde{\gamma}$,$\tilde{Z}$ \\ \vspace{-0.4cm} \\ \\ \\  \\
          $\tilde{\gamma}$,$\tilde{Z}$,$\tilde{H}$ \\ \\
          $\tilde{H}$ \\ \vspace{-0.6cm} \\ \\ \\
       \end{tabular}
     & \begin{tabular}{ccccll}
         $\tilde{q}$ & $\longrightarrow$
                     & $q$ & $\chi_1^0$ &  & \\
         &  &        & $\stackrel{\lambda'}{\hookrightarrow}$
                     & $\nu \bar{q}' q'$   &  \\
         $\tilde{u}_L$ & $\longrightarrow$
                     & $d$ & $\chi_1^+$   &  & \\
         &  &        & $\hookrightarrow$
                     & $W^+$ & $\hspace{-1.1cm}\chi_1^0$ \\
         &  &  & & $\:|$
         & $\hspace{-1.1cm}
           \stackrel{\lambda'}{\hookrightarrow}$ $\nu \bar{q}' q'$
           \vspace{-0.2cm} \\
         &  &  & & $\:\mid$ \vspace{-0.3cm} &  \\
         &  &  &     & $\:\rightarrow$  $q \:\: \bar{q}'$ & \\
         $\tilde{u}_L$ & $\longrightarrow$ & $d$ & $\chi_1^+$ &  & \\
         &  &        & $\stackrel{\lambda'}{\hookrightarrow}$
                     & $\nu u \bar{d}$ &  \\
         $\tilde{u}_L$ & $\longrightarrow$
                     & $d$ & $\chi_1^+$ &  & \\
         &  &        & $\hookrightarrow$
                     & $W^+$ & $\hspace{-1.1cm}\chi_1^0$ \\
         &  &  &     & $\hookrightarrow$ $q \:\: \bar{q}'$ &
       \end{tabular}
     & \begin{tabular}{l}
        Missing $P_T$ \\
        + multiple jets
       \end{tabular}\\                                          \hline
  S6 & \begin{tabular}{c}
          $\tilde{H}$ \\ \vspace{-0.1cm} \\ \\
       \end{tabular}
     & \begin{tabular}{ccccll}
         $\tilde{u}_L$ & $\longrightarrow$
                     & $d$ & $\chi_1^+$ &  & \\
         &  &        & $\hookrightarrow$
                     & $W^+$ & $\hspace{-1.1cm}\chi_1^0$ \\
         &  &  &     & $\hookrightarrow$  $l^+ \:\: \nu$  &
       \end{tabular}
     &
      \begin{tabular}{l}
         High $P_T$ $e^+$ or $\mu^+$ \\
         + missing $P_T$ + 1 jet
       \end{tabular}\\                                          \hline
  S7 & \begin{tabular}{c}
          $\tilde{\gamma}$,$\tilde{Z}$\\ \vspace{-0.5cm} \\ \\ \\ \\
       \end{tabular}
     & \begin{tabular}{ccccll}
         $\tilde{u}_L$ & $\longrightarrow$
                     & $d$ & $\chi_1^+$  &  & \\
         &  &        & $\hookrightarrow$
                     & $W^+$ & $\hspace{-1.1cm}\chi_1^0$ \\
         &  &  &  & $\:|$
         & $\hspace{-1.1cm}
           \stackrel{\lambda'}{\hookrightarrow}$ $e^{\pm} \bar{q}' q''$
           \vspace{-0.2cm} \\
         &  &  &     & $\:\mid$ \vspace{-0.3cm} &  \\
         &  &  &     & $\:\rightarrow$  $l^+ \:\: \nu$  &
       \end{tabular}
     & \begin{tabular}{l}
        High $P_T$ $e^{\pm}$ \\
        + high $P_T$ $e^+$ or $\mu^+$ \\
        + missing $P_T$ \\
        + multiple jets
       \end{tabular}\\                                          \hline
  S8 & \begin{tabular}{c}
          $\tilde{\gamma}$,$\tilde{Z}$\\ \vspace{-0.5cm} \\ \\ \\ \\
       \end{tabular}
     & \begin{tabular}{ccccll}
         $\tilde{u}_L$ & $\longrightarrow$
                     & $d$  & $\chi_1^+$  &  & \\
         &  &        & $\hookrightarrow$
                     & $W^+$ & $\hspace{-1.1cm}\chi_1^0$ \\
         &  &  &     & $\:|$
         & $\hspace{-1.1cm}
           \stackrel{\lambda'}{\hookrightarrow}$ $\nu \bar{q}' q'$
           \vspace{-0.2cm} \\
         &  &  &     & $\:\mid$ \vspace{-0.3cm} &  \\
         &  &  &     & $\:\rightarrow$  $l^+ \:\: \nu$   &
       \end{tabular}
     & \begin{tabular}{l}
        High $P_T$ $e^+$ or $\mu^+$ \\
        + missing $P_T$ \\
        + multiple jets
       \end{tabular}\\
   \hline \hline
  \end{tabular}
  \caption[]
          {\small \label{tab:sqtopo2}
               Squark decay channels in \Rp\ SUSY classified per
               distinguishable event topologies (second part).
               As in table~\ref{tab:sqtopo1}, the list of processes
               given here is not exhaustive, e.g. the gauge decays
               $\chi_1^+ \rightarrow \chi_1^0 l^+ \nu$ and
               $\chi_1^+ \rightarrow \chi_1^0 q \bar{q}'$
               may also proceed via a virtual sfermion. }
 \end{center}
 
\end{table}
 
 
The possible decay modes of the chargino, when it is the lightest
chargino $\chi_1^+$, are the gauge decays
$\chi_1^+ \rightarrow \chi_1^0 l^+ \nu$ and
$\chi_1^+ \rightarrow \chi_1^0 q \bar{q}'$,
and the \Rp\ decays $\chi_1^+ \rightarrow \nu u \bar{d}$ and
$\chi_1^+ \rightarrow e^+ d \bar{d}$.
The fate of the $\chi_1^0$ depends on its gaugino-higgsino
composition.
The question of how this $\chi_1^0$ nature depends on free fundamental
parameters of the MSSM, as well as the corresponding $\tilde{q}$
branching fractions for various possible decay channels will be
discussed briefly in relation to our analysis in
section~\ref{sec:results} and was studied in more detail
in~\cite{SUSY95,DRPEREZ}.
In general, the $\chi_1^0$ will undergo the decay
$\chi_1^0 \rightarrow e^{\pm} q \bar{q}'$ or
$\chi_1^0 \rightarrow \nu q \bar{q}$.
The former will be dominant if the $\chi_1^0$ is photino-like
(i.e. dominated by photino components) in which case both the
``right'' and the ``wrong'' sign lepton (compared to incident
beam) are equally probable leading to largely background free
striking signatures for lepton number violation.
The latter will dominate if the $\chi_1^0$ is zino-like.
A higgsino-like $\chi_1^0$ will most probably
be long lived and escape detection since its coupling to
fermion-sfermion pairs (e.g. Fig.~\ref{fig:sqdiag}d) is proportional
to the fermion mass~\cite{GUNION}.
Hence processes involving a $\tilde{H}$-like $\chi_1^0$ will be
affected by an imbalance in transverse momenta.
 
 
Taking into account the dependence on the nature of the $\chi_1^0$,
the possible decay chains of the $\tilde{u}_L$ and $\bar{\tilde{d}}_R$
squarks can be classified into eight distinguishable event
topologies listed in tables~\ref{tab:sqtopo1} and~\ref{tab:sqtopo2}
and labelled {\large S1} to {\large S8}.
 
 
For a squark decaying into a quark and the lightest neutralino,
the partial width can be written as
$$ \Gamma_{\tilde{q}\rightarrow \chi_1^0 q}
 = \frac{1}{8\pi} \left( A^2 + B^2 \right) M_{\tilde{q}}
   \left(1-\frac{M^2_{\chi_1^0}}{M^2_{\tilde{q}}}\right)^2
\Longrightarrow
   \Gamma_{\tilde{q}\rightarrow \tilde{\gamma}q}
 = \Gamma_{\tilde{q}\rightarrow eq'} \; \frac{2e^2e^2_q}{\lambda'^2} \;
   \left(1-\frac{M^2_{\tilde{\gamma}}}{M^2_{\tilde{q}}}\right)^2  $$
where $A$ and $B$ in the left expression are chiral couplings
depending on the mixing parameters.
Detailed expressions for such couplings can be found in~\cite{GUNION}.
Under the simplifying assumption that the neutralino is a pure
photino $\tilde{\gamma}$, this gauge decay width reduces to the
expression on the right. Here we introduced the partial width
$ \Gamma_{\tilde{q}\rightarrow eq'} = \lambda'^2 M_{\tilde{q}}/16 \pi$
for squarks undergoing \Rp\ decays.
It is seen that, in general, gauge decays contribute strongly at low
$\chi_1^0$ masses and small Yukawa couplings.
 
\begin{figure*}[htb]
 \begin{center}
  \begin{tabular}{p{0.40\textwidth}p{0.60\textwidth}}
   \vspace{-5.0cm}
   \caption[]{ \label{fig:bgfstop}
     {\small Stop pair production via $\gamma$-gluon fusion at HERA,
             followed by \Rp\ decay of the $\tilde{t_{1}}$.}} &
     \mbox{\epsfxsize=0.60\textwidth \epsffile{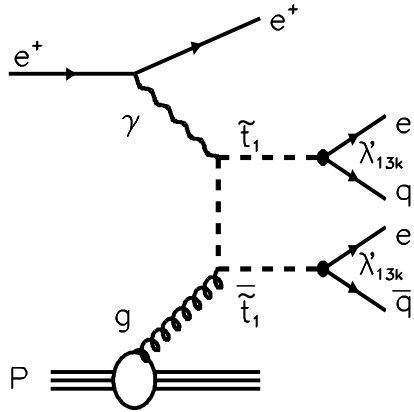}}
 \end{tabular}
 \end{center}
 \vspace{-1.0cm}
\end{figure*}

 
The case $\lambda'_{131} \ne 0 $ (or $\lambda'_{132} \ne 0$) is of
special interest~\cite{KONRP} since it allows for direct production
of the stop via
$e^+ d \rightarrow \tilde{t}$ ($e^+ s \rightarrow \tilde{t}$).
The stop is particular in the sense that a ``light'' stop
mass eigenstate ($\tilde t_{1}$) could
(depending upon the mass parameters for the chiral states
and on the free parameters of the model)
exist much lighter than the top quark itself and lighter than other
squarks.
This applies only for the stop since the off-diagonal terms which
appear in the mass matrix associated to the superpartners
of chiral fermions are proportional to the partner fermion mass.
Such a stop $\tilde{t}_1$ mass eigenstate is considered in this paper
and its search is furthermore extended towards low mass by considering
pair production via $\gamma$-gluon fusion as illustrated in
Fig.~\ref{fig:bgfstop}.
%
%
For the study of this process, we assume that the $\tilde{t}_1$
is lighter than the lightest chargino.
Hence the $\tilde{t}_1$ will decay dominantly into a positron and
a quark since, by assumption, the decays into $t \chi_1^0$ and
$b \chi_1^+$ are forbidden and the one-loop decay into
$c \chi_1^0$ is negligible even for small values of the \Rp\
coupling of the $\tilde{t}_1$ to a positron-quark pair~\cite{SUSY95}.
 
\section{The H1 detector}
\label{sec:h1det}
 
A detailed description of the H1 detector can be found
in~\cite{H1DETECT}.
Here we describe only the components relevant for the present analysis
in which the event final state involves either an $e^+$ (or $e^-$)
with high transverse energy or a large amount of hadronic transverse
energy flow.
 
The $e^+$ (or $e^-$) energy and angle are measured in a finely
segmented liquid argon (LAr) sampling calorimeter~\cite{H1LARCAL}
covering the polar
angle\footnote{The incoming proton moves by definition in the forward
               ($z > 0$) direction with $\theta=0^{\circ}$ polar angle.}
range 4$^{\circ} \le \theta \le$ 153$^{\circ}$ and all azimuthal angles.
It consists of a lead/argon electromagnetic section
followed by a stainless-steel/argon hadronic section.
Electromagnetic energies are measured with a resolution
of $\sigma(E)/E \simeq$ $12$ \%/$\sqrt{E}\oplus1\%$ and hadronic
energies with $\sigma(E)/E \simeq$ $50$ \%/$\sqrt{E}\oplus2\%$
after software energy weighting~\cite{H1CALRES}.
The absolute scales are known to 2\% and 5\% for
electromagnetic and hadronic energies respectively.
The angular resolution on the scattered electron measured from the
electromagnetic shower in the calorimeter is $\lesssim 4$ mrad.
A lead/scintillator electromagnetic backward calorimeter extends the
coverage at larger angles
(155$^{\circ} \le \theta \le$ 176$^{\circ}$).
 
Located inside the calorimeters is the tracking system used here to
determine the interaction vertex and the charge of the final state
lepton.
The main components of this system are central drift and proportional
chambers (25$^{\circ} \le \theta \le$ 155$^{\circ}$), a forward track
detector  (7$^{\circ} \le \theta \le$ 25$^{\circ}$) and backward
proportional chambers (155$^{\circ} \le \theta \le$ 175$^{\circ}$).
The tracking chambers and calorimeters are surrounded
by a superconducting solenoid coil providing a uniform field of
$1.15${\hbox{$\;\hbox{\rm T}$}}
within the tracking volume.
The instrumented iron return yoke surrounding this coil is used to
measure leakage of hadronic showers and to recognize muons.
The luminosity is determined from the rate of the Bethe-Heitler process
$e p \rightarrow e p \gamma$ measured in a luminosity monitor.
 
\section{Analysis}
\label{sec:analyz}
 
\subsection{Single production of squarks}
 
\label{sec:single}
 
 
For the search for resonant production of squarks, the event selection
basically relies on the final state lepton finding and on global
energy-momentum conservation cuts.
It is optimized separately for each of the event topologies
(see tables~\ref{tab:sqtopo1} and~\ref{tab:sqtopo2})
{\large S1} to {\large S8} by relying on Monte Carlo simulation.
 
 
The simulation of the leptoquark-like signatures ({\large S1} and
{\large S2}) relies on the event generator LEGO~\cite{LEGOSUSS}.
For squarks undergoing gauge decays followed by a $\chi_1^0$ or
$\chi_1^+$ \Rp\ decay into a high $P_T$ $e^{\pm}$ and multiple
jets, i.e. processes belonging to topologies {\large S3} and
{\large S4}, the generator SUSSEX~\cite{LEGOSUSS} based on the
cross-sections given in~\cite{RPVIOLATION} is used.
Both generators also simulate initial state bremsstrahlung
in the collinear approximation, initial and
final state parton showers and fragmentation~\cite{PYTHIA,JETSET73},
and properly take into account the correction of the kinematics
at the decay vertex for effects of the parton shower masses.
The parton densities used~\cite{MRSDM,MRSHSF} are evaluated at the
scale of the new particle mass, and this scale is also chosen for the
maximum virtuality of parton showers.
For these channels, a complete simulation of the H1 detector response
is performed.
The event topologies {\large S5} to {\large S8} (as well as some of the
processes in {\large S3} or {\large S4} which proceed through
the exchange of a virtual $W$ or virtual sfermion),
were studied at four-vector level~\cite{DRPEREZ} taking into account
matrix element calculations~\cite{BARTL} and multiparticle phase space.
For these channels, realistic efficiencies are then obtained by
smearing the particle four-vectors according to measured resolutions,
detector effects and acceptances.
The efficiencies thus obtained were cross-checked and found to agree
typically within $5\%$ with a complete simulation based on SUSSEX
for those {\large S5} processes where the $\chi_1^+$ undergoes a \Rp\
violating decay.
 
 
A complete Monte Carlo simulation of the H1 detector response is
performed for each possible background source.
For the DIS NC or CC background estimates we make use of either
the DJANGO~\cite{DJANGO} or the LEPTO~\cite{INGELMAN} event
generator.
DJANGO includes first order radiative corrections and simulation
of real bremsstrahlung photons based on HERACLES~\cite{HERACLES},
as well as QCD dipole parton showers based on ARIADNE~\cite{ARIADNE}.
LEPTO includes the lowest order electroweak scattering process
with QCD corrections to first order in $\alpha_s$, complemented by
leading-log parton showers and string fragmentation~\cite{JETSET74}.
Both generators agree in channels where one expects
a single hard jet, i.e. {\large S1}, {\large S2} and {\large S6}.
The LEPTO event generator is used in the multijet channels
{\large S3}, {\large S4}, {\large S5}, {\large S7} and {\large S8}.
 
The parton densities in the proton used for DIS throughout are taken
from the MRS~H~\cite{MRSHSF} parametrization which is close to
$F_2$ structure function measurements at HERA (see~\cite{HERASF}).
For the direct and resolved photoproduction of light and heavy flavours,
the PYTHIA MC event generator~\cite{PYTHIA} is used which includes QCD
corrections to first order in $\alpha_s$, leading-log parton showers
and string fragmentation~\cite{JETSET73}.
The GRV LO (GRV-G LO) parton densities~\cite{SFGRVGLO} in the proton
(photon) are used at low $Q^2$.
 
 
The event selection for real data starts with the rejection of
non-colliding background. This selection step is common to all channels
and requires: \\
\begin{enumerate}
 
  \item a primary interaction vertex in the range
        $\mid z - \bar{z} \mid < 35 \cm$ with $\bar{z} = 3.4 \cm$;
  \item that the event survives a set of halo and cosmic muon filters;
        for channel {\large{S2}} these are complemented by visual scan;
  \item that the event be properly in time relative to interacting bunch
        crossings.
 
\end{enumerate}
Cut (1) mainly suppresses beam--wall, beam--residual gas and, with (2)
and (3), background from cosmic rays and halo muons.
We moreover impose that the events be accepted by LAr calorimetry
triggers~\cite{H1LARCAL}:
the events of {\large{S1}}, {\large{S3}} and {\large{S4}} must satisfy
``electron'' or ``transverse energy'' trigger requirements;
events of {\large{S2}} and {\large{S5}} must fulfill
``missing transverse energy'' requirements;
events of {\large{S6}}, {\large{S7}} and {\large{S8}} must satisfy
either ``electron'' or ``missing transverse energy'' requirements.
 
The selection cuts and data reduction specific to each of the event
topologies for the $ep$-induced background is presented below.
In each case the number of event candidates observed are compared to
SM expectations. The systematic (syst.) errors quoted on the mean
expected background in each case take into account uncertainties on
the absolute electromagnetic and hadronic energy scales (see
section ~\ref{sec:h1det}), on the integrated luminosity ($1.5\%$)
and the contribution due to finite Monte Carlo statistics.
Estimates of SUSY signal detection efficiencies are also given in
each channel.
 
 
\hfill \\
\begin{flushleft}
{\bf Event topology {\large S1}:}
\end{flushleft}
 
\noindent
For the event topology {\large S1}, i.e. events characterized by
the DIS NC-like signatures~\cite{H1LQ95}, it is necessary to
reject contaminating background from other physical processes.
We require:
\begin{enumerate}
  \item an isolated `$e^+$' cluster~\cite{H1CALEPI}
        with $E_{T,e} = E_e \sin \theta_e > 7 \GeV$
        and $10^{\circ} \le \theta_e \le  145^{\circ}$;
       (here `$e^+$' includes all $e^{\pm}$ candidates except
        those having an associated track with explicitly measured
        negative charge);
        the isolation requires that less than 10\% additional energy
        be found within a pseudorapidity-azimuth cone of opening
        $ \sqrt{ (\Delta \eta_e)^2 + (\Delta \phi_e)^2 } < 0.25 $
        centered on the $e^{\pm}$ candidate;
  \item that if two `$e^+$' cluster candidates are found, they must not
        be balanced in $E_{T,e}$ and in azimuth,
        i.e. $E^1_{T,e} / E^2_{T,e} > 1.25$
        and $ \mid \Delta \phi_{1,2} - 180^{\circ} \mid > 2^{\circ} $,
        and the candidate with highest $E_{T,e}$ must be at smallest
        rapidity;
  \item a total missing transverse momentum
        $P_{T,miss} \approx
        \sqrt{\left(\sum E_x^i \right)^2 + \left(\sum E_y^i \right)^2}
        \leq 15 \GeV$  summed over all energy depositions $i$ in the
        calorimeters, with $ E_x^i = E^i \sin \theta^i \cos \phi^i $
        and $ E_y^i = E^i \sin \theta^i \sin \phi^i $;
  \item a minimal ``longitudinal momentum'' loss in the direction of
        the incident positron,
        $ -8 \leq 2 E_e^0 - \sum \left(E - P_z\right) \leq 12 \GeV$,
        where $E_e^0$ is the incident positron beam energy;
  \item a $y_e$, measured from the final state `$e^+$',
        satisfying $ y_e < 0.95 $.
\end{enumerate}
Cuts (1) and (3) eliminate DIS CC events.
Cut (2) suppresses QED Compton events.
Cut (4) provides a powerful rejection of photoproduction contamination
and also suppresses DIS NC-like events with a very hard $\gamma$
emitted from the initial state positron.
Cut (5) further suppresses photoproduction with a ``fake'' $e$ which
tends to cluster at largest $y_e$ for largest $M_e$, where $y_e$
is the standard DIS Lorentz invariant and $M_e$ the ``squark mass''
reconstructed from the final state `$e^+$' energy $E_e$ and
angle $\theta_e$ as :
$$ M_e = \sqrt{ s x_e } = \sqrt{\frac{Q^2_e}{y_e}}, \;\;\;\;
     Q^2_e = \frac{E^2_{T,e}}{1-y_e}, \;\;\;\;
       y_e = 1 - \frac{E_e - E_e \cos \theta_e}{2E_e^0}; \;\;\;\;  $$
$Q^2$ is the standard momentum transfer squared of DIS.
In addition to the above requirements, we apply a $M_e$ dependent $y_e$
cut which is designed~\cite{H1LQ94,H1LQ95} via Monte Carlo studies to
optimize the signal significance for scalar leptoquark searches,
given the expected background.
This $y_e$ cut varies from $y_e > 0.5$ at $45 \GeV$ to $y_e > 0.35$ at
$150 \GeV$ and down to $y_e > 0.05$ at $275 \GeV$.
 
\begin{figure}[htb]
  \begin{center}
    \begin{tabular}{cc}
      \mbox{\epsfxsize=8.0cm \epsffile{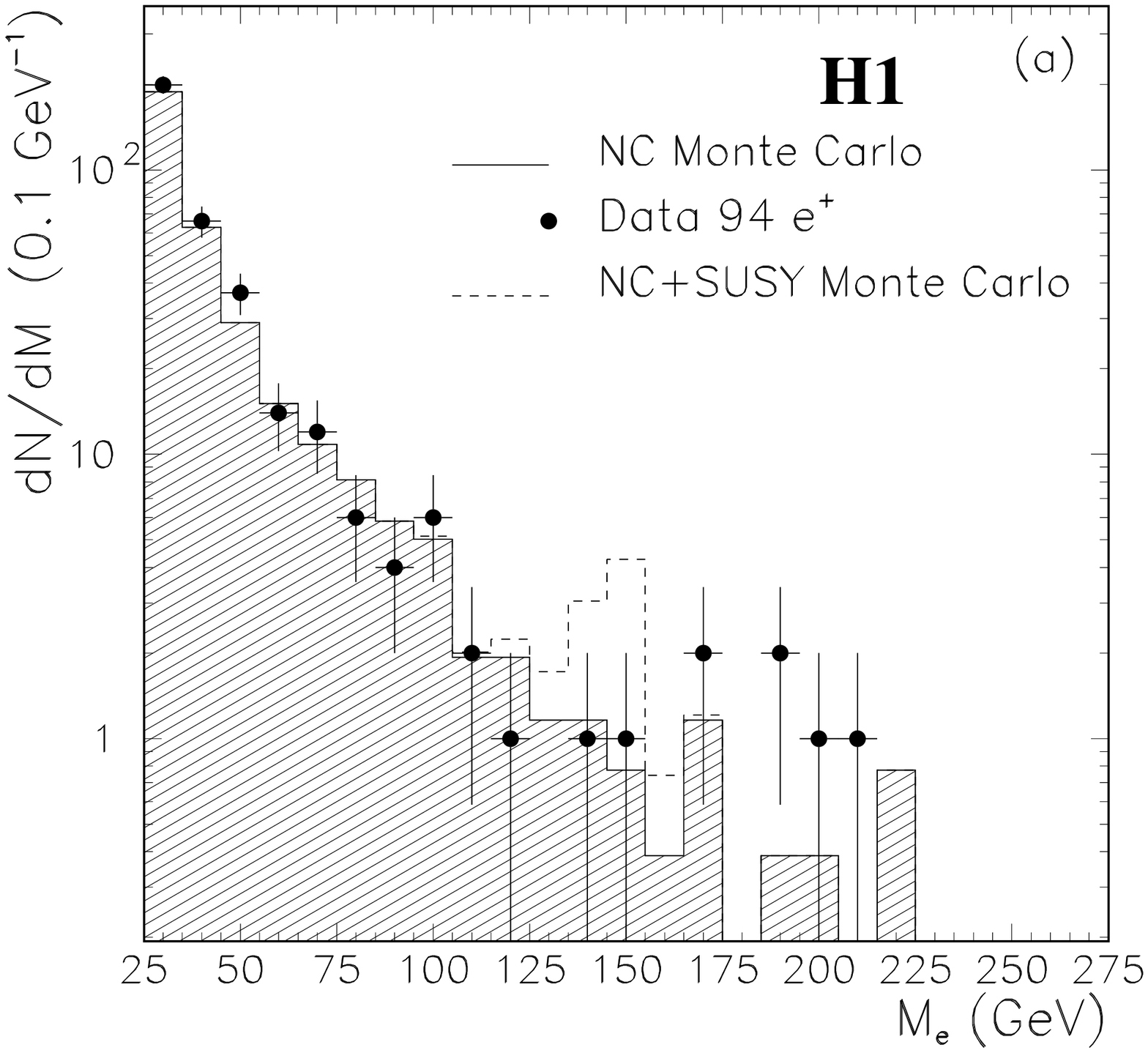}}
      \mbox{\epsfxsize=8.0cm \epsffile{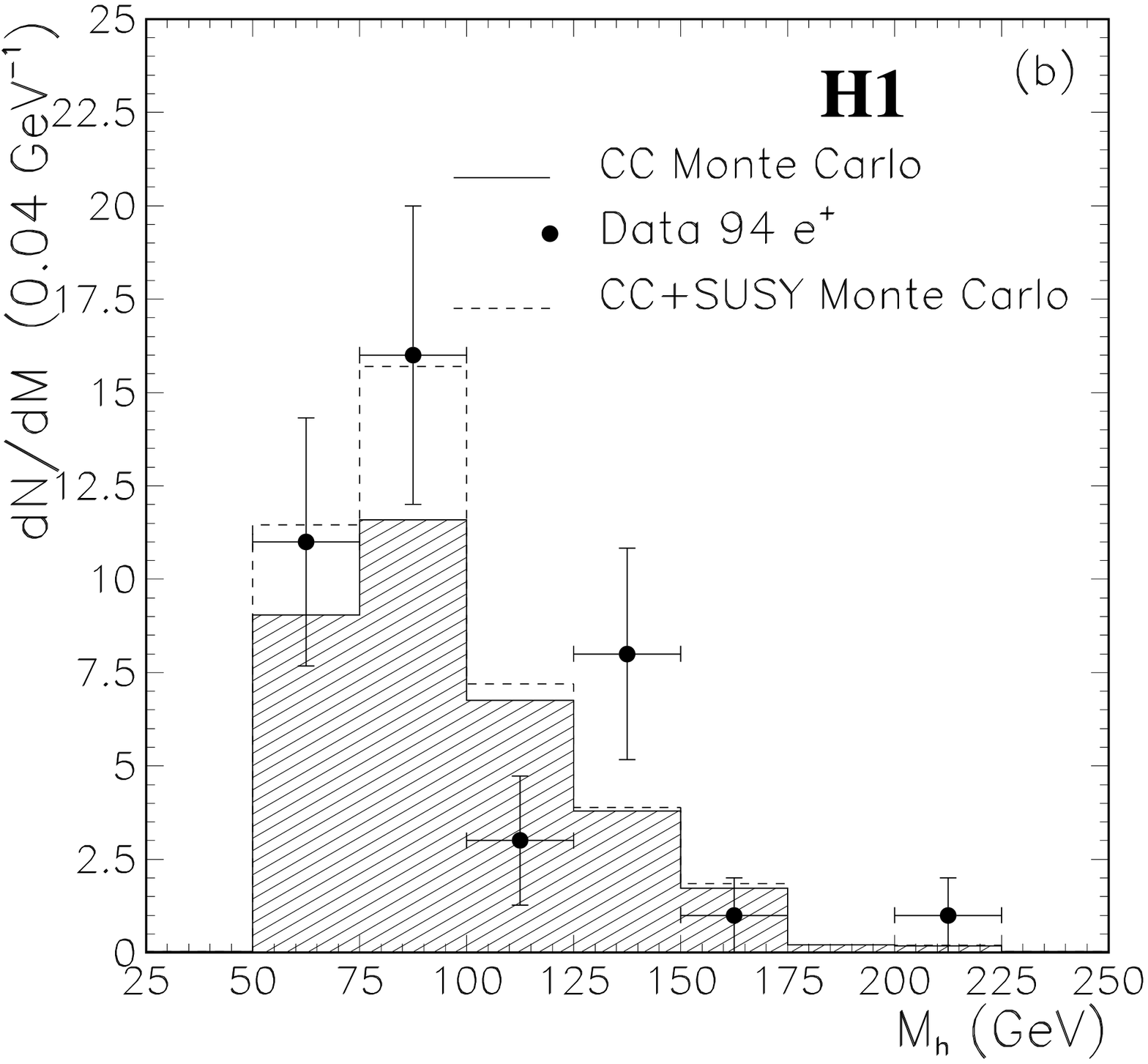}}
    \end{tabular}
  \end{center}
\caption[]{ \label{fig:dndmnccc}
  {\small Mass spectra for (a) $e + q$ (b) $\nu + q$ final states
       for data (closed points) and DIS Monte Carlo (shaded
       histograms). The superimposed dashed histograms show typical \Rp\
       SUSY signals near the sensitivity limit
       (see table~\ref{tab:sqresu} in section~\ref{sec:results})
       for (a)
       $M_{\tilde q} = 150 \GeV$ and $M_{\chi_1^0} = 20 \GeV$
       and for (b)
       $M_{\tilde q} = 75 \GeV$ and $M_{\chi_1^0} = 20 \GeV$. }}
\end{figure}

For these NC-like (leptoquark like) signatures, 362 events
satisfy the selection requirements and the $y_e$ cut in the mass range
$M_e > 25 \GeV$.
This observed number of events is in good agreement with the mean
expected DIS NC background of $335 \pm 36$ (syst.) events.
The measured mass spectrum is compared to the DIS NC expectation in
Fig.~\ref{fig:dndmnccc}.
%
For $ M_e > 45 \GeV$, we are left with 91 events while
$84 \pm 10.2$ (syst.) events are expected from the SM.
For $ M_e > 100 \GeV$, 13 events are observed in good agreement
with the mean SM expectation of $12.4 \pm 2.6$ (syst.).
 
In this channel, the \Rp\ SUSY signal detection efficiency is found
to be weakly dependent on $M_{\tilde{q}}$ and ranges from
$43\%$ at $45 \GeV$ to $68\%$ at $150 \GeV$ in the
middle of the mass range considered here.

 
\hfill \\
\begin{flushleft}
{\bf Event topology {\large S2}:}
\end{flushleft}
 
\noindent
The event topology {\large S2} is characterized by DIS CC-like
signatures~\cite{H1LQ95} for which we require:
\begin{enumerate}
  \item no $e^{\pm}$ cluster satisfying the above {\large S1}
        requirements;
  \item $P_{T,miss} > 25 \GeV$;
  \item the total transverse energy
        $E_T \approx \sum \mid \vec{P}_T \mid$
        calculated from energy depositions in the calorimeter
        should match the total missing transverse momentum
        $P_{T,miss}$ such that $(E_T - P_{T,miss})/E_T < 0.5$.
\end{enumerate}
Cuts (1) to (3) eliminate photoproduction and DIS NC
background.
 
In total, 40 CC-like events satisfy all above requirements in the
relevant mass and $y$ range at $ M_h > 45 \GeV$ and
$ y_h < 0.95 $ where $M_h$ and $y_h$ are reconstructed by
summing over all measured final state hadronic energy:
$$ M_h = \sqrt{\frac{Q^2_h}{y_h}}, \;\;\;\;
     Q^2_h= \frac{P^2_{T,miss}}{1-y_h},\;\;\;\;
       y_h=\frac{\sum \left(E-P_z\right)}{2E_e^0}.\;\;\;\; $$
As is seen in Fig.~\ref{fig:dndmnccc}b, this is in good agreement
with the DIS CC expectation of $33.4 \pm 3.6$ (syst.) events.
 
The \Rp\ SUSY signal detection efficiency in this channel rises from
$\sim 15\%$ at $45 \GeV$ to reach a plateau at $\sim 80\%$ above
$100 \GeV$.

 
\hfill \\
\begin{flushleft}
{\bf Event topology {\large S3}:}
\end{flushleft}
 
\begin{figure}[htb]
  \begin{center}
    \begin{tabular}{cc}
         \mbox{\epsfxsize=8.0cm \epsffile{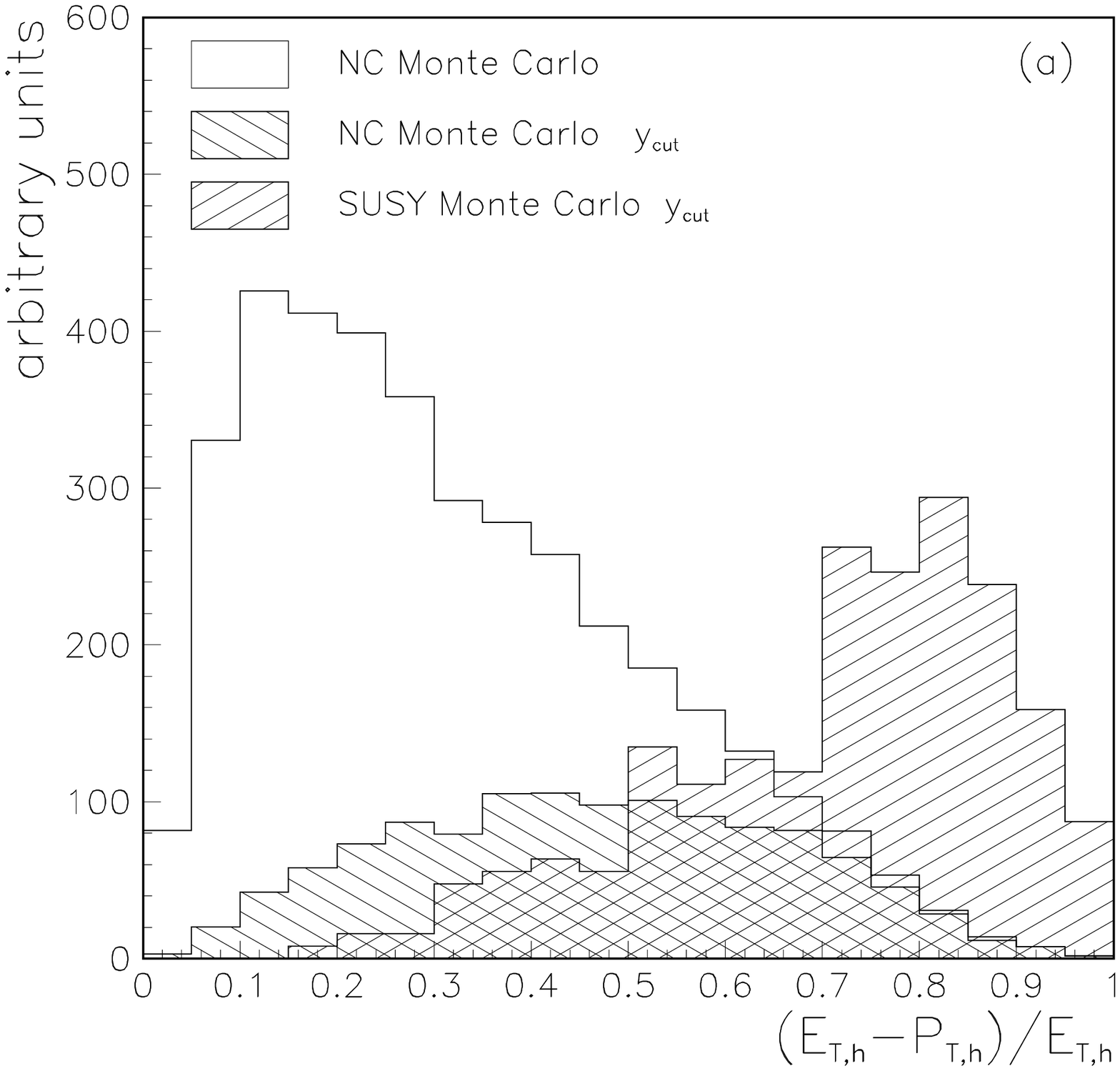}}
         \mbox{\epsfxsize=8.0cm \epsffile{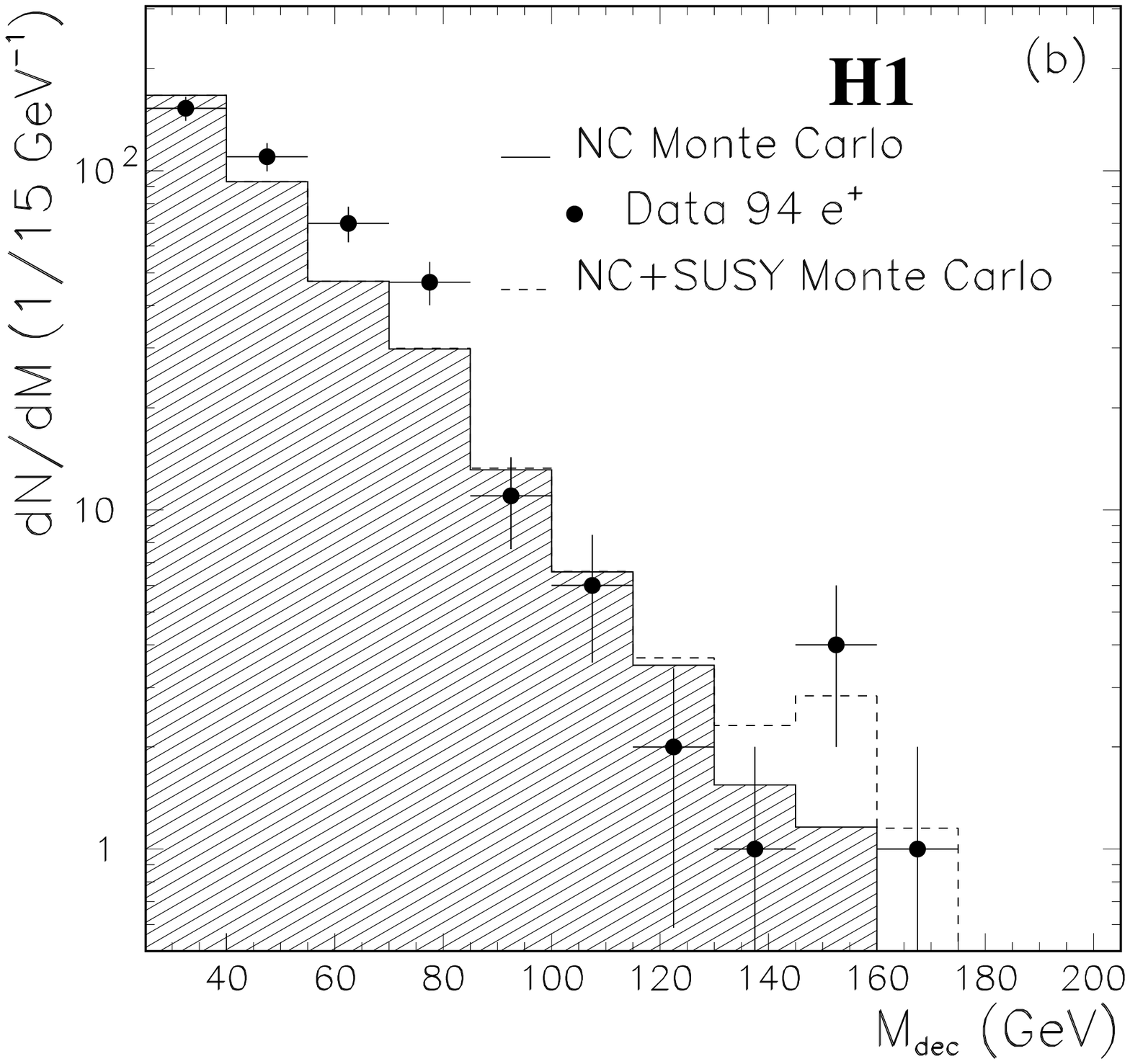}}
    \end{tabular}
  \end{center}
  \caption[]{ \label{fig:dndmsu94}
    {\small (a) Balance between the total scalar hadronic energy
         $E_{T,h}$ and the vector sum $P_{T,h}$ for DIS NC before and
         after the $y_{e} > 0.4$ cut and for a simulation of a $150 \GeV$
         squark decaying into $e^+ q \rightarrow e^+ q'' \bar{q''} q'$;
         (b) measured mass spectrum (uncorrected) for the squark
         candidates in the $e^+ q \rightarrow e^+ q'' \bar{q''} q'$
         channel for data (dots) and DIS NC Monte Carlo (histogram),
         the superimposed dashed histogram in (b) show a typical \Rp\
         SUSY signal near the sensitivity limit
        (see table~\ref{tab:sqresu} in section~\ref{sec:results}) for
         $M_{\tilde q} = 150 \GeV$ and $M_{\chi_1^0} = 40 \GeV$.}}
\end{figure}

\noindent
For a gauge decay of the squarks leading to a ``right''
sign final state lepton  (i.e. $ e^{+} q \rightarrow \tilde{q}
                                  \rightarrow \chi_1^0 + q'
                                  \rightarrow e^{+} q'' \bar{q}'' q'$),
we impose the following stringent requirements in complement to
{\large{S1}} cuts :
\begin{enumerate}
   \item the `$e^+$' must give $y_e > 0.4$;
   \item an imbalance between the total hadronic $E_{T,h}$ and
         $P_{T,h}$ such that $(E_{T,h} - P_{T,h}) / E_{T,h} >0.25$;
   \item at least one reconstructed jet with $P_{T,jet} > 7 \GeV$;
   \item the azimuthal opening angle $\Delta \phi_{jt}$ between the
         jet of highest $P_T$ (i.e. generally the current jet in a
         DIS NC process) and the axis defined by the total hadronic
         transverse momentum
           $\vec{P}_{T,h} = (\sum E_{x,h}, \sum E_{y,h})$
         be larger than
         $ \Delta \phi_{jt} > 2/5 \times (50 - E_{T,h})$ with
         $E_{T,h}$ in $\GeV$ and $\Delta \phi_{jt}$ in degrees;
   \item the squark invariant mass calculated from all final state
         particles ($M_{dec}$) excluding the proton
         fragments~\cite{H1LQ94} must deviate from $M_e$ by more
         than $10\%$.
\end{enumerate}
Cut (1) strongly suppresses DIS NC background.
The $y_e$ distribution calculated from the `$e^+$' in SUSY events
appears strongly shifted towards large $y_e$ since the
squark decays uniformly in its center-of-mass frame and, further,
since the `$e^+$' takes away only a fraction of the
$\chi_1^0$ momentum.
Cut (2) exploits the fact that the hadronic energy of the event is
not concentrated within one single jet because of the decay products
of the $\chi_1^0$. Fig.~\ref{fig:dndmsu94}a shows how this value
discriminates the DIS NC background from the signal.
The jet finding for cuts (3) and (4) relies (here and throughout the
paper) on a simple cone algorithm in the laboratory reference frame
with a fixed pseudorapidity-azimuth opening radius of 1 unit.
Cut (4) further suppresses lowest order DIS NC events by imposing
sufficient hadronic activity far enough in azimuth from the ``current''
jet.
Cut (5) ensures that the events accepted here (gauge decay modes) are
not simultaneously accepted as {\large S1} candidates (\Rp\ decay modes).
%
 
We find 405 candidates satisfying the previous cuts for masses
$M_{dec} > 25 \GeV$, which is to be compared with the mean SM background
of $363 \pm 39$ events expected from DIS NC.
The measured mass spectrum is compared to Monte Carlo expectations
in Fig.~\ref{fig:dndmsu94}b.
Above $45 \GeV$ we observe 220 events in the data, while
$154 \pm 17$ (syst.) events are expected from DIS NC.
This represents an excess of 2.9 standard deviations in the Gaussian
limit approximation (combining statistical and systematic errors in
quadrature).
The slight excess of events is seen to be mostly concentrated at low
masses and in particular in the mass range from $40$ to $85 \GeV$.
Nevertheless, it should be recalled here
that our DIS NC Monte Carlo for multijet channels (LEPTO) does not
include full QED corrections which could lead to a migration of events
with true $y_e$ below cut (1) towards larger apparent $y_e$.
A good agreement is observed for $M_{dec} > 100 \GeV$ where we
find $14$ events while the mean DIS NC expectation is
$13.2 \pm 2.6$ (syst.) events.
For $M_{dec} > 140 \GeV$, 5 events are found which agrees well
with the expectation of $1.9 \pm 0.9$ (syst.).
 
In this channel, the \Rp\ SUSY signal detection efficiencies (which
experimentally sums that of both the $ right \, sign$ and the
$ unsigned $ events), depend mainly on the $\chi_1^0$.
For $M_{\chi_1^0} = 20 \GeV$ it rises from $\sim 20\%$ for
$M_{\tilde{q}} = 45 \GeV$ to a plateau of $\sim 33\%$ for
$M_{\tilde{q}} \gtrsim 75 \GeV$.
For $M_{\chi_1^0} = 80 \GeV$ it reaches $\sim 60\%$ for
$M_{\tilde{q}} \gtrsim 100 \GeV$.

 
\hfill \\
\begin{flushleft}
{\bf Event topology {\large S4}:}
\end{flushleft}
 
\noindent
For a gauge decay of the squarks leading to a ``wrong'' sign
final state lepton  (i.e. $ e^{+} q \rightarrow \tilde{q}
                            \rightarrow e^{-} q'' \bar{q}'' q'$) we
perform a determination of the lepton charge
using the tracking chamber information.
Hence, we impose in addition to the above {\large S3} criteria :
\begin{enumerate}
  \item the $e^-$ LAr cluster must be geometrically linked to a
        negatively charged track and the cluster energy must match
        the track momentum within $ \mid (E - P)/( E + P) \mid < 0.5 $;
  \item the track must be made of at least 40 digitisations in the
        central tracking chamber;
  \item the error in the curvature $\kappa$ must fulfil
         $ \mid \kappa / \delta \kappa \mid > 1 $.
\end{enumerate}
These cuts ensure a good quality of the track reconstruction and
track-cluster matching at the expense of a reduced efficiency (partly
due to occasional inoperation of either the inner or the outer central
drift chambers) for accepting the $e^-$ track of about $70\%$ in
the angular range $\theta \ge 35^{\circ}$ well covered by the central
tracking chambers.
 
We observe no $e^-$ (wrong sign) events among the 405 candidates
satisfying the kinematical requirements for squark gauge decays.
 
In this channel, the charge track requirements imply an additional
efficiency loss compared to {\large S3} for the \Rp\ SUSY signal which
is negligible at $45 \GeV$ but which increases to $10\%$ at $150 \GeV$
and $20\%$ at $250 \GeV$.
 
 
\hfill \\
\begin{flushleft}
{\bf Event topology {\large S5}:}
\end{flushleft}
 
\noindent
The event topology {\large S5} is characterized by large missing
transverse momentum and multiple jets. We require:
\begin{enumerate}
  \item no $e^{\pm}$ cluster satisfying the above {\large S1}
        requirements;
  \item $P_{T,miss} > 15 \GeV$;
  \item $(E_{T,h} - P_{T,h})/E_{T,h} > 0.25$;
  \item $ P_{T,h} > 50 \times (1 - (E_{T,h} - P_{T,h})/E_{T,h}) $
        with $P_{T,h}$ in $\GeV$;
   \item at least one reconstructed jet with $P_{T,jet} > 7 \GeV$;
         the jet of highest $P_T$ should satisfy
         $ \Delta \phi_{jt} > (4/7) \times (100 - E_{T,h}) $
         with $\Delta \phi_{jt}$ in degrees and $E_{T,h}$ in $\GeV$.
\end{enumerate}
Cut (3) exploits the fact that more than one jet is expected in such
events. Cut (4) removes the DIS CC background, which is mainly
concentrated at low values of $(E_{T,h}-P_{T,h})/E_{T,h}$. Cut (5)
removes photoproduction events for which one of the two back-to-back
jets is badly measured, so that the $ \Delta \phi_{jt}$ is expected
to be small.
We are left with $9$ events in the data sample compared to an expectation
of $3.9 \pm 4$ (syst.) events from $\gamma p$ photoproduction background
and a negligible DIS CC background.
Here a sizeable contribution to the systematic error originates
from the dependence of the LAr trigger efficiency on this
{\large S5} multijet topology.
 
\begin{figure}[htb]
  \begin{center}
 
    \vspace{-1.0cm}
 
    \begin{tabular}{cc}
      \mbox{\epsfxsize=7.3cm \epsffile{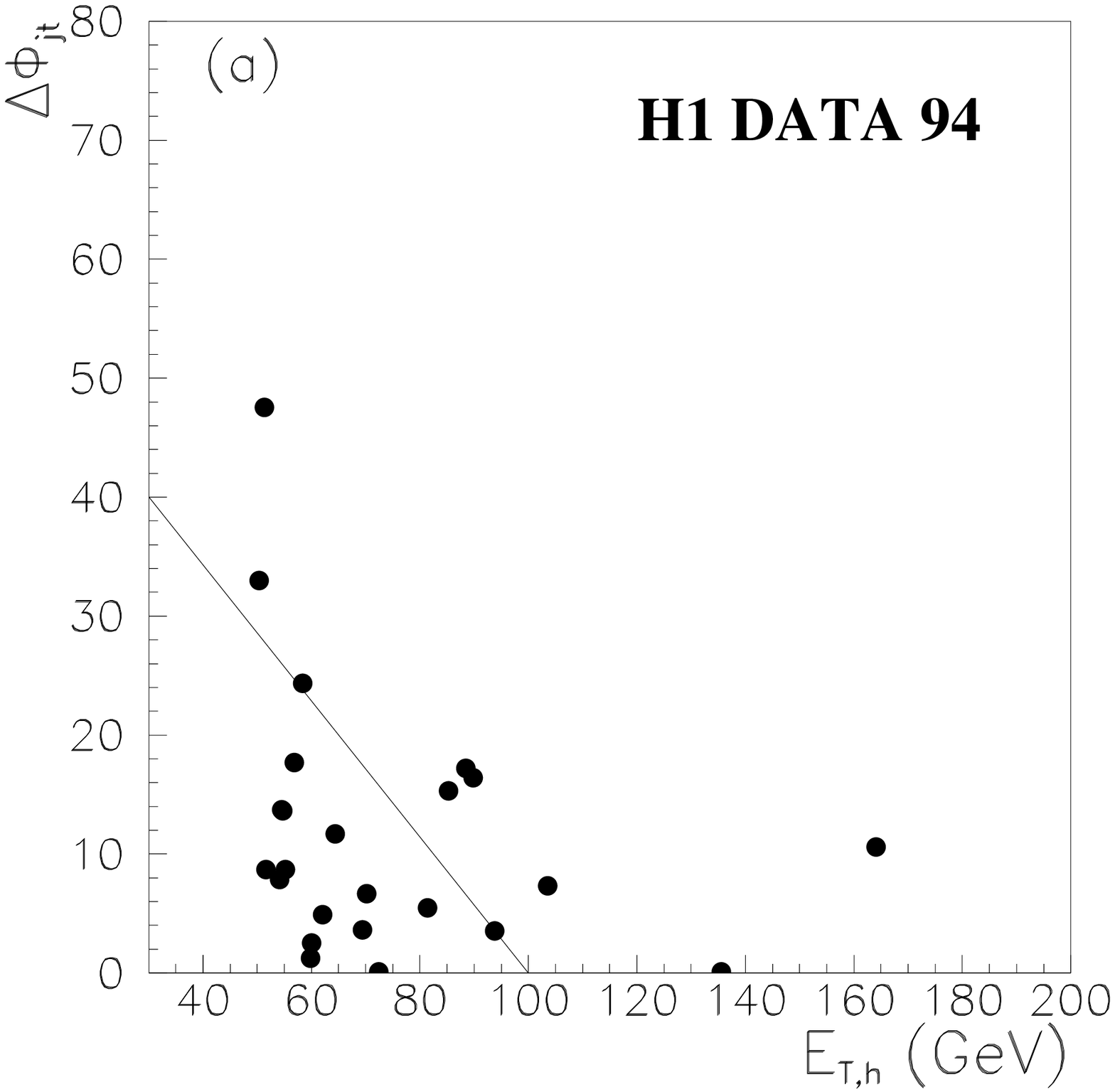}}
      \mbox{\epsfxsize=7.3cm \epsffile{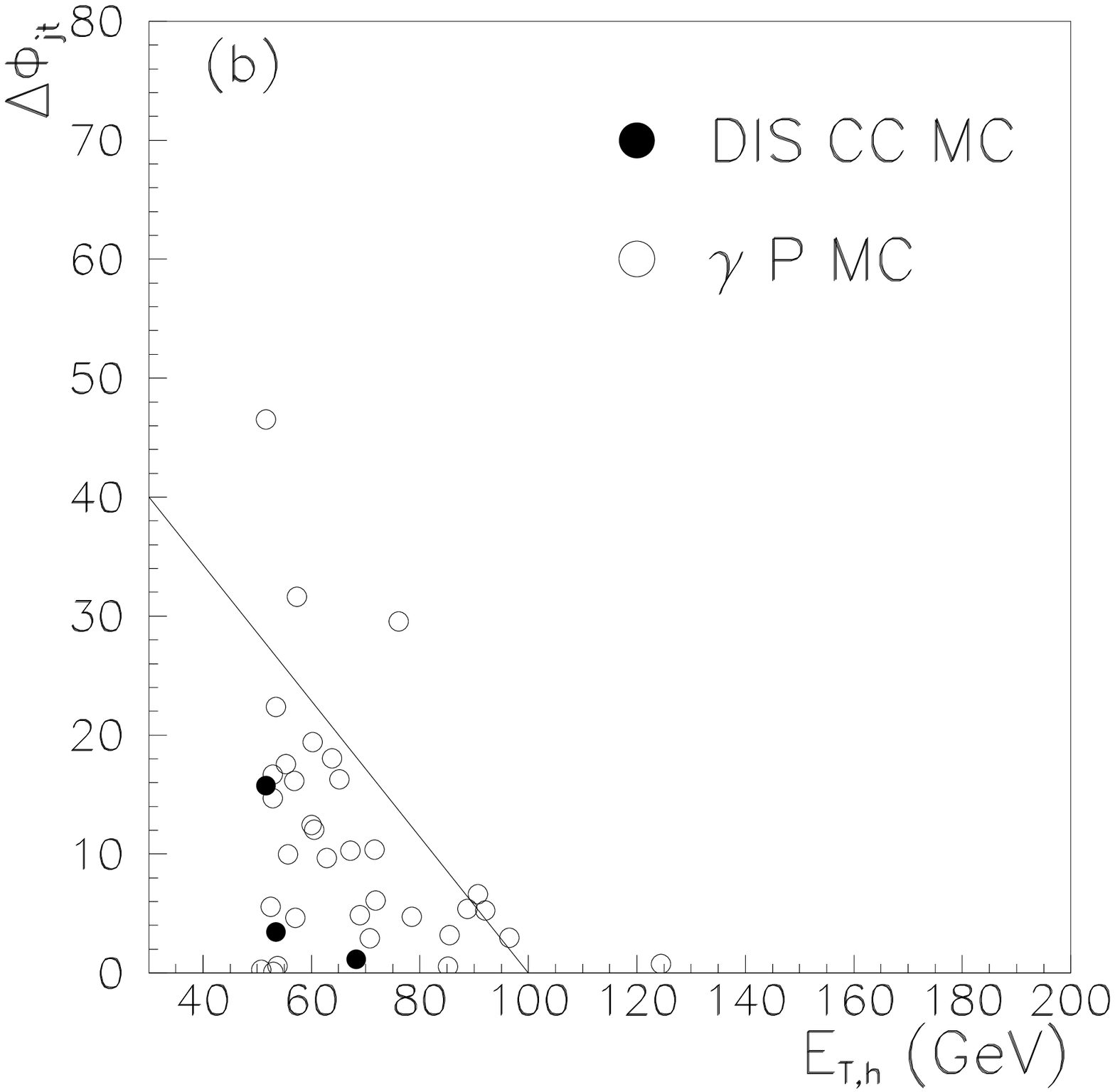}}
    \end{tabular}
 
    \vspace{-0.5cm}
 
    \begin{tabular}{cc}
      \mbox{\epsfxsize=7.3cm \epsffile{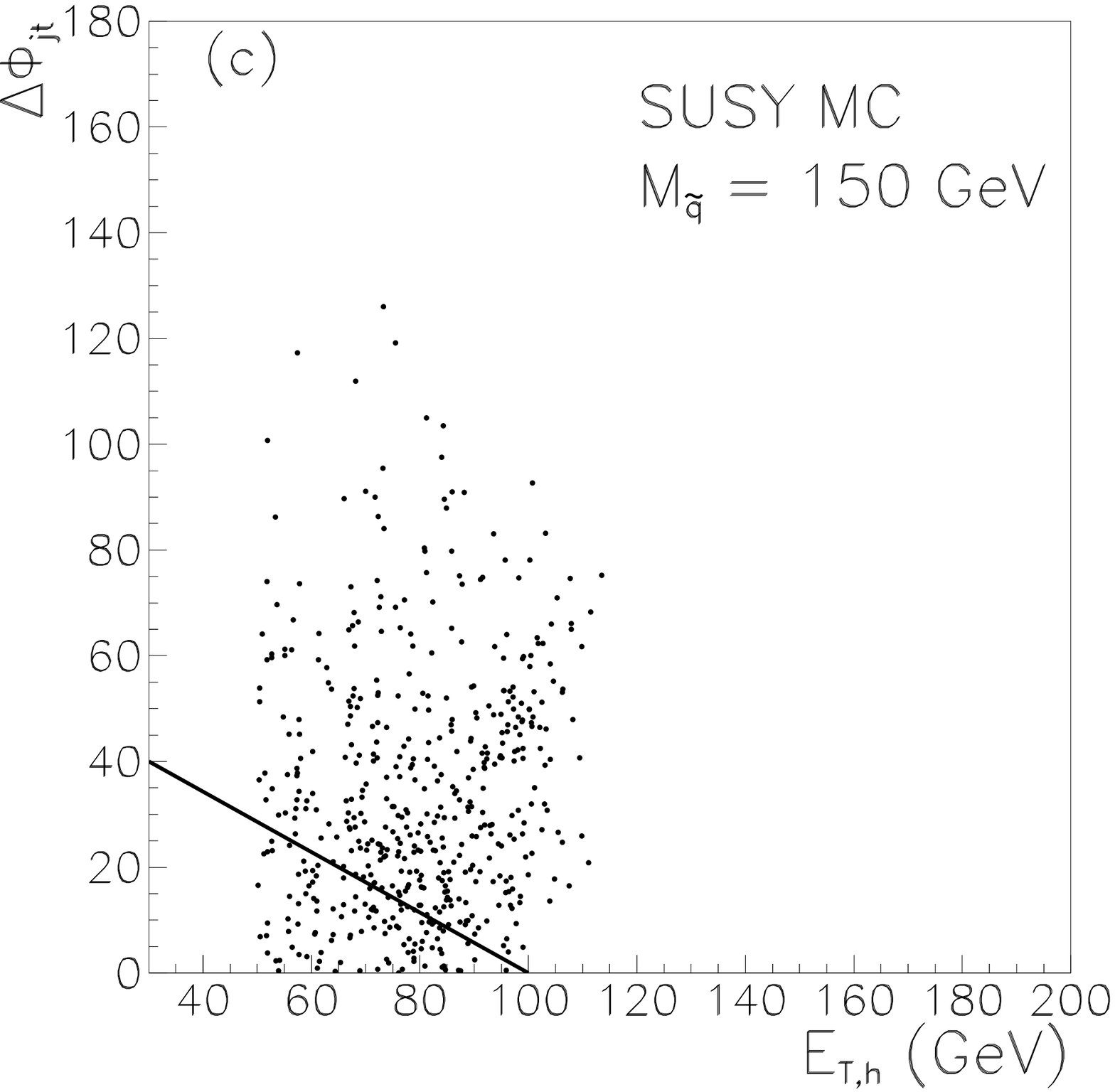}}
      \mbox{\epsfxsize=7.3cm \epsffile{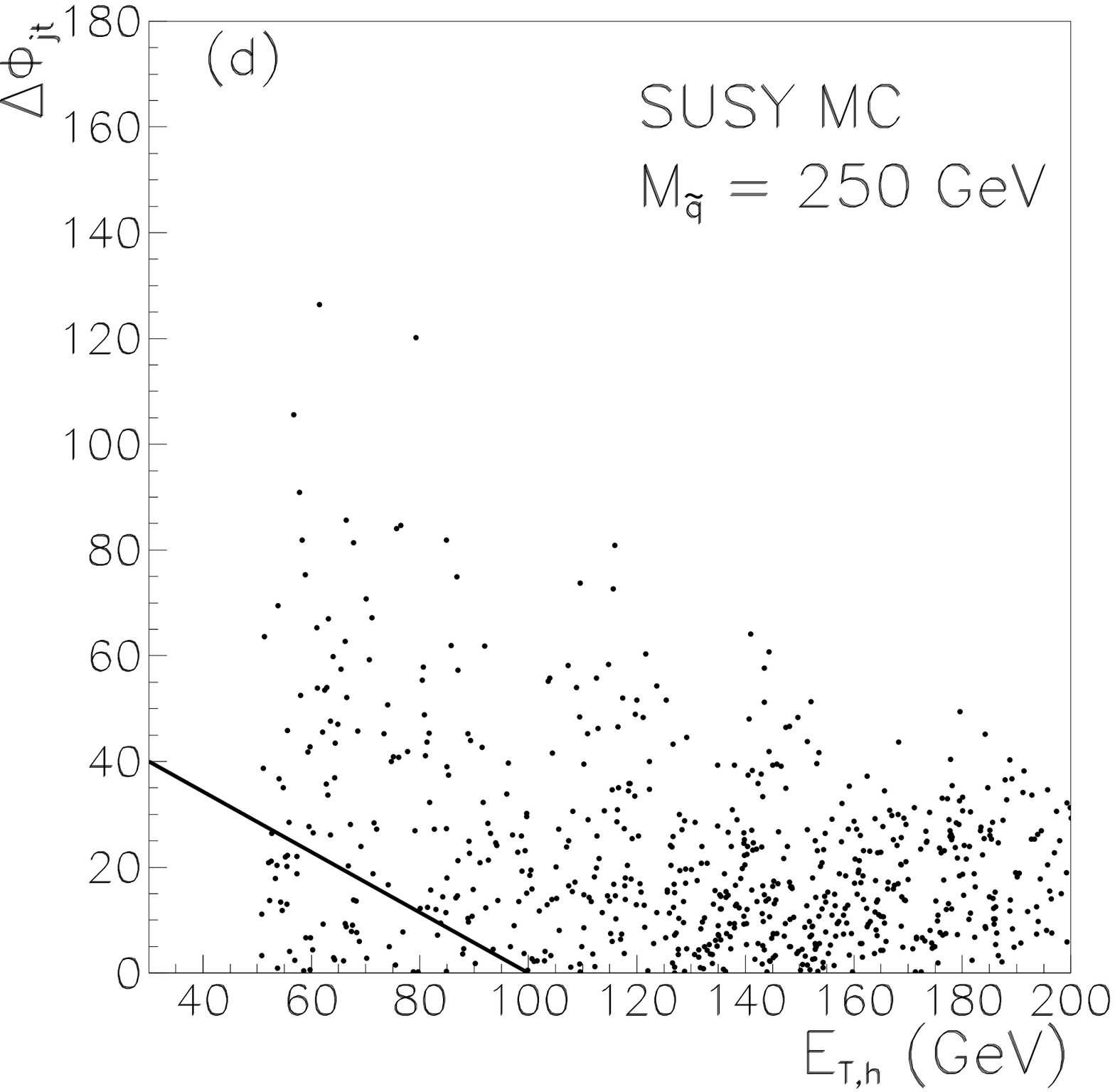}}
    \end{tabular}
 \end{center}
 
\vspace{-0.5cm}
 
\caption[]{ \label{fig:dphicut}
  {\small  Correlation between the azimuthal opening angle
           $\Delta \phi_{jt}$ and the total hadronic transverse
           energy $E_{T,h}$ for (a) data, (b) DIS CC and
           $\gamma p$ background MC and (c), (d) \Rp\ SUSY signal
           in topology {\large S5} for two example cases.
           The events above the cut (solid line) are accepted. }}
\end{figure}

It is shown in Fig.~\ref{fig:dphicut} how these 9 remaining events
compare to the SM photoproduction and DIS CC expectations (respectively
for $1 \times {\cal{L}}_{data}$ and $10 \times {\cal{L}}_{data}$).
No LAr trigger efficiency losses are folded in the Monte Carlo sample of
Figs.~\ref{fig:dphicut}b,c, and d.
It can be seen also that cut (5) still ensures a good efficiency
for a possible SUSY signal.
%
 
The efficiency for \Rp\ SUSY events in {\large S5} rises with increasing
$M_{\tilde{q}}$ up to a plateau for $M_{\tilde{q}} \gtrsim 150 \GeV$.
It also rises with increasing $M_{\chi_1^0}$ mainly because of the
$P_{T,miss}$ selection cut imposed.
For processes where the final state $\chi_1^0$ (or $\chi_1^+$)
at the squark decay vertex directly undergoes a \Rp\ decay, the
efficiency at $M_{\tilde{q}} \gtrsim 150 \GeV$ is $\sim 26\%$ for
$M_{\chi_1^0} = 20 \GeV$ and $\sim 52 \%$ for $M_{\chi_1^0} = 80 \GeV$.
When the $\chi_1^0$ is $\tilde{H}$-like (stable) the mass difference
between the $\chi_1^0$ and the $\chi_1^+$ is relatively smaller for
heavier $\chi_1^0$ and this hampers the signal detection.
Hence in processes where the $\chi_1^0$ escapes detection, the
efficiency which at $M_{\tilde{q}} \gtrsim 150 \GeV$ reaches
$\sim 50\%$ for $M_{\chi_1^0} = 20 \GeV$ is down to $\sim 30\%$ for
$M_{\chi_1^0} = 80 \GeV$.
Finally, processes where the $\chi_1^+$ undergoes a cascade decay
(e.g.  $\chi_1^+ \rightarrow W^+ \chi_1^0        \; ; \;
        \chi_1^0 \rightarrow e^+ \bar{q}' q''   \; ; \;
         W^+  \rightarrow    q \bar{q}'$),
suffers from an efficiency loss due to the $P_{T,miss}$ cut.
In such case the efficiency rises from $\sim 6\%$ at
$M_{\tilde{q}} \sim 100 \GeV$ to $\sim 32 \%$ at
$M_{\tilde{q}} \gtrsim 175 \GeV$.
 
 
\hfill \\
\begin{flushleft}
{\bf Event topology {\large S6}:}
\end{flushleft}
 
\noindent
The event topology {\large S6} is characterized by the presence
of a lepton ($e^+$ or $\mu^+$) at large transverse energy $E_{T,l}$,
a large missing transverse momentum and a single jet.
 
\noindent
To search for cases where the final state lepton is a positron, we
require:
\begin{enumerate}
  \item an isolated `$e^+$' with $E_{T,e} > 7 \GeV$;
  \item $P_{T,miss} > 15 \GeV$;
  \item $(E_{T,h} - P_{T,h})/E_{T,h} < 0.5$;
  \item $ 0.4 < y_e < 0.95 $;
  \item at least one reconstructed jet with $P_{T,jet} > 7 \GeV$;
  \item at least 1 charged track with $P_{track} > 5 \GeV$ originating
        from the primary vertex and linked to the electron cluster.
\end{enumerate}
Cut (1) suppresses DIS CC background while cut (2) suppresses
photoproduction background. The other cuts are designed to optimize
the specific {\large S6} signal significance relative to tails of
background distributions.
We are left with $2$ event candidates in the data while
$3.8 \pm 1.3$ (syst.) are expected from DIS NC background.
 
\begin{figure}[htb]
 \vspace{-1.0cm}
 
 \begin{flushleft}
 
  \epsfig{file=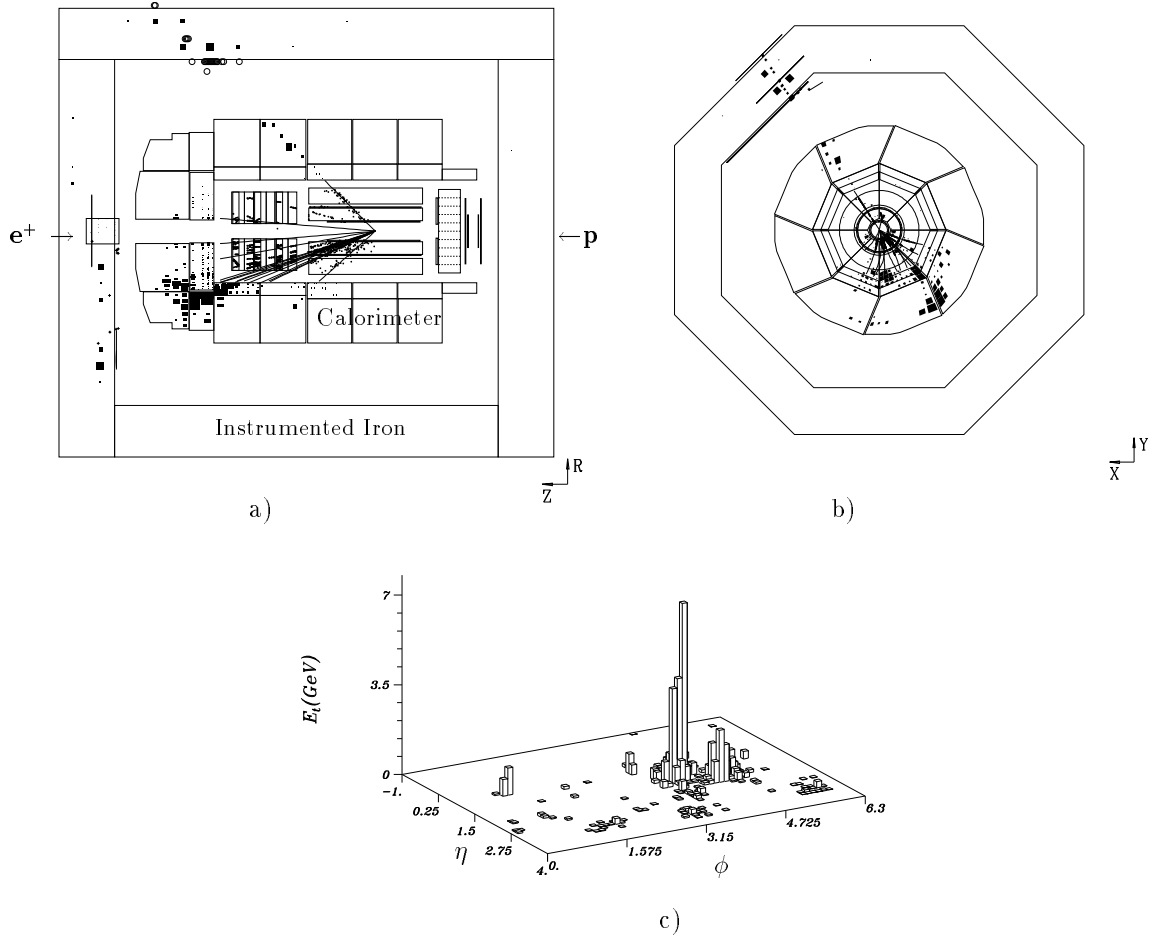,bbllx=50pt,%
                            bblly=450pt,%
                            bburx=550pt,%
                            bbury=800pt,%
                            width=16.0cm,%
                            angle=0}
 
 \end{flushleft}
 \vspace{2.0cm}
 
\caption[]{ \label{fig:thevent}
  {\small Display of the $e^+ p \rightarrow \mu+ \; jet$ candidate
          from H1 run number 84295 showing (a) an R-$z$ view,
          (b) a R-$\phi$ view and (c) the calorimetric transverse
          energy flow. }}
\end{figure}

\noindent
To search for cases where the final state lepton is a muon, we
require:
\begin{enumerate}
  \item no isolated $e^{\pm}$ with $E_{T,e} > 7 \GeV$;
  \item $P_{T,miss} > 25 \GeV$ calculated from the energy depositions in
        the calorimeters;
  \item $(E_{T,h} - P_{T,h})/E_{T,h} < 0.5$;
  \item at least one reconstructed jet with $P_{T,jet} > 7 \GeV$;
  \item at least 1 charged track with
        $P_{track} > 10 \GeV$ linked to the primary vertex and
        lying outside of a $\Delta \phi = 60^{\circ}$ cone
        centered on the direction of the jet at highest $P_{T,jet}$;
  \item that the charged track in cut (5) be ``penetrating'' in the sense
        that there be $ < 5 \GeV$ of total energy measured in the LAr
        calorimeter within a $\Delta \phi = 15^{\circ}$ cone
        centered on the track direction;
        moreover the track should not point to a
        cluster localized around the azimuthal cracks in the LAr
        calorimeter.
\end{enumerate}
We are left with $1$ event candidate in the data of which an event
display is shown Fig.~\ref{fig:thevent}.
A remarkable ``$\mu \; + \; jet$'' signature is seen
(Fig.~\ref{fig:thevent}a) with a positively charged isolated track.
%
A detailed analysis of this event~\cite{MUEVENT} reveals that the
isolated track has a transverse momentum of
$23 \pm 2.4^{+7}_{-5} \GeV$
and appears in azimuth at $\Delta \phi_{\mu,h} = 183 \pm 1^{\circ}$
from an hadronic system (Fig.~\ref{fig:thevent}b)
which itself has in total $P_{T,h} = 42.1 \pm 4.2 \GeV$.
This hadronic system is built from two main ``clusters''
(Fig.~\ref{fig:thevent}c) which are found to be merged into a single
jet with our cone algorithm within a radius of 1 unit in the
pseudorapidity-azimuth plane.
This leaves overall a total missing transverse momentum of
$ \mid \vec{P}_{T,h} + \vec{P}_{T,\mu} \mid
                                        = 18.7 \pm 4.8^{+5}_{-7} \GeV $
and a ``longitudinal momentum'' loss in the incident positron
direction of
$ 2 E_e^0 - \{ \sum \left( E_h - P_{z,h} \right) +
                    \left( E_{\mu} - P_{z,\mu} \right) \} =
                                       35.8 \pm 1.6^{+3}_{-2.1} \GeV $.
 
With our selection cuts (1) $\rightarrow$ (6), we expect on average
from DIS CC a background of $ 0.04 \pm 0.03 $ (syst.) event
corresponding to a probability of about $4\%$ to observe here one
or more such event.
The actual dominating background is expected~\cite{MUEVENT}
to come from associated $W^+$ production in NC and CC processes
followed by a leptonic decay of the $W^+$,
$ e^+ p \rightarrow e^+ W^+ X \rightarrow e^+ \mu^+ \nu X$,
where the final state $e^+$ is lost in the beam pipe.
With our selection cuts, this process studied in a Monte Carlo
calculation using the SM cross-section of~\cite{BAUR} gives a $15\%$
probability for such a background event to be accepted.
 
The efficiency for \Rp\ SUSY events in {\large S6} rises with increasing
$M_{\tilde{q}}$ up to a plateau for $M_{\tilde{q}} \gtrsim 150 \GeV$, and
decreases with increasing $M_{\chi_1^0}$ for reasons already explained
in topology {\large S5} for $\tilde{H}$-like $\chi_1^0$.
For $M_{\tilde{q}} \gtrsim 150 \GeV$ it is of $\sim 50\%$ for
$M_{\chi_1^0} = 20 \GeV$, and down to $\sim 35\%$ for
$M_{\chi_1^0} = 80 \GeV$.
 
It is interesting to note~\cite{KONKOKI} that an \Rp\ SUSY signal
is consistent with the properties of the observed event candidate.
Imposing more restrictive requirements such as:
$ 2 E_e^0 - \{ \sum \left( E_h - P_{z,h} \right) +
                    \left( E_{\mu} - P_{z,\mu} \right) \} > 12 \GeV$,
$\Delta \phi_{\mu,h} > 140^{\circ}$,
$P_{T,h} > 40 \GeV$ and  $P_{T,\mu} > 10 \GeV$,
we find that more than $30\%$ of the \Rp\ SUSY events satisfying the
selection cuts (1) to (6) in the muon channel also verify these
additional requirements provided that the $\tilde{u}_L$ be in the mass
range $100 \lesssim M_{\tilde{u}_L} \lesssim 200 \GeV$ and the
$\chi_1^+$ be in the range
$20 \lesssim M_{\chi_1^+} \lesssim 30 \GeV$ for
                               $M_{\tilde{u}_L} \simeq 100 \GeV$
or
$20 \lesssim M_{\chi_1^+} \lesssim 110 \GeV$ for
                               $M_{\tilde{u}_L} \simeq 200 \GeV$.
 
With the above stringent requirement that $P_{T,h} > 40 \GeV$
applied on the background, which implies in associated
$W^+$ production a very stiff recoiling hadronic system,
one is left with a $\sim 3\%$ probability for such interpretation of
the event and a negligible contribution from misidentified DIS CC events.
 
 
\hfill \\
\begin{flushleft}
{\bf Event topology {\large S7}:}
\end{flushleft}
 
\noindent
The event topology {\large S7} is characterized by the presence
of an $e^{\pm}$ at large $E_T$, accompanied by another lepton $l^{+}$,
large missing transverse momentum and multiple jets.
We require:
\begin{enumerate}
  \item an isolated $e^{\pm}$ with $E_{T,e} > 7 \GeV$ and giving
        $0.4 < y_e < 0.95$;
  \item $P_{T,miss} > 15 \GeV$;
  \item $(E_{T,h} - P_{T,h})/E_{T,h} > 0.25$;
  \item at least one reconstructed jet with $P_{T,jet} > 7 \GeV$.
\end{enumerate}
In the case where $l^{+}$ is a positron we require in addition
a second isolated $e^{\pm}$ with $E_T > 5 \GeV$.
We observe no candidate in the data while $0.4 \pm 0.3$ event is
expected on average from DIS NC and none from DIS CC.
\noindent
In the case where $l^{+}$ is a muon we impose
a more stringent cut on $P_{T,miss} > 25 \GeV$.
We observe no candidate in the data, while $0.4 \pm 0.3$ event is
expected on average from DIS NC and none from DIS CC.
 
The efficiency for \Rp\ SUSY events in {\large S7} depends weakly on
$M_{\tilde{q}}$ and rises with  $M_{\chi_1^0}$.
It is of $\sim 20\%$ for $M_{\chi_1^0} = 20 \GeV$ and rises in the case
where $l^+$ is a positron to $\sim 60\%$ for $M_{\chi_1^0} = 80 \GeV$.
In the case where $l^+$ is a muon and for heavy $\chi_1^0$,
the apparent $P_{T,miss}$ tends to be reduced by the quasi-collinearity
of the $\mu$ and neutrino in the final state and the efficiency only
reaches $\sim 30\%$ for $M_{\chi_1^0} = 80 \GeV$.
 
 
\hfill \\
\begin{flushleft}
{\bf Event topology {\large S8}:}
\end{flushleft}
 
\noindent
The event topology {\large S8} is characterized by the presence of
one charged lepton ($e^+$ or $\mu^+$) at large $E_T$, a large missing
transverse momentum and multiple jets.
 
For the case where the lepton is a positron, we require:
\begin{enumerate}
  \item an isolated `$e^+$' with $E_{T,e} > 7 \GeV$ and giving
        $ 0.4 < y_e < 0.95 $;
  \item $P_{T,miss} > 15 \GeV$;
  \item $(E_{T,h} - P_{T,h})/E_{T,h} > 0.25$;
  \item at least one reconstructed jet with $P_{T,jet} > 7 \GeV$.
\end{enumerate}
We are left with 3 candidates in the data while $2.3 \pm 1.0$ are
expected from DIS NC.
The background from DIS CC and photoproduction is here negligible.
 
For the case where the lepton is a muon, we require:
\begin{enumerate}
  \item no isolated $e^{\pm}$ with $E_{T,e} > 7 \GeV$;
  \item $P_{T,miss} > 25 \GeV$;
  \item $(E_{T,h} - P_{T,h})/E_{T,h} > 0.25$;
  \item at least one reconstructed jet with $P_{T,jet} > 7 \GeV$;
  \item at least 1 ``penetrating'' and isolated charged track with
        $P_{track} > 10 \GeV$, such tracks were defined in cuts (5) and
        (6) of the {\large S6} selection.
\end{enumerate}
We are left in the data with the same $\mu+ X$ event candidate as was
found in {\large S6} event topology while
$0.04 \pm 0.03$ event is expected from DIS CC background.
 
The efficiency for \Rp\ SUSY events in {\large S8} has similar
$M_{\tilde{q}}$ and $M_{\chi_1^0}$ dependence as in {\large S7}.
It is of $\sim 20-30\%$ for $M_{\chi_1^0} = 20 \GeV$
and rises in the case where $l^+$ is a positron
to $\sim 65\%$ for $M_{\chi_1^0} = 80 \GeV$.
In the case where $l^+$ is a muon, it remains
at $\sim 25\%$ for $M_{\chi_1^0} = 80 \GeV$.
 
While the efficiency is rather high for the \Rp\ SUSY signal to satisfy
the basic cuts (1) to (5), it is here (contrary to the {\large S6}
channel) difficult to meet the more restrictive
conditions~\cite{KONKOKI}
that could be imposed on the observed $\mu^+ X$ event candidate.
In particular,
the overlap of hadronic $P_T$ flow initiated by the three quark jets
into a single observed jet and with $\Delta \phi_{\mu,h} > 140^{\circ}$
is a unlikely configuration.
It is moreover difficult to satisfy simultaneously the stringent cut
$ 2 E_e^0 - \{ \sum \left( E_h - P_{z,h} \right) +
                    \left( E_{\mu} - P_{z,\mu} \right) \} > 12 \GeV$.
We find that less than $10\%$ of the \Rp\ SUSY events satisfying the
selection cuts (1) to (5) also verify these additional requirements
for $\tilde{u}_L$ be in the mass range
$100 \lesssim M_{\tilde{u}_L} \lesssim 200 \GeV$.
 
\subsection{Pair production of stop squarks}
 
For squarks which are pair produced in $\gamma$-gluon fusion
processes (Fig.~\ref{fig:bgfstop}), the scattered electron is generally
lost in the beam pipe. The optimization of the event selection relies
on Monte Carlo simulation.
The simulation of stop pair production in $\gamma$-gluon
fusion~\cite{DRPEREZ} is based on the $\sigma_{stop}$ cross-section
calculated in the Weizs\"acker-Williams approximation.
The resolved photon contribution~\cite{DREES} as well as the
contribution from a $Z_0$ boson exchange are neglected.
The GRV LO gluon density in the proton~\cite{SFGRVGLO} is used.
The background simulation is based on the event generator
DJANGO~\cite{DJANGO} for DIS NC and PYTHIA~\cite{PYTHIA} for
photoproduction processes.
We impose:
\begin{enumerate}
  \item $P_{T,miss} \le 15 \GeV$;
  \item two $e^{\pm}$ candidates $i$ satisfying $E^i_{T,e} > 5 \GeV$
        within $10^{\circ} \le \theta^i_e \le  145^{\circ}$,
        the $e^{\pm}$ must be isolated within pseudorapidity-azimuth
        cones of opening
        $ \sqrt{ (\Delta \eta^i_e)^2 + (\Delta \phi^i_e)^2 } < R_i $
        where $R_{i=1} = 0.5$ and $R_{i=2} = 0.25$;
  \item the two $e^{\pm}$ candidates must be acollinear in the transverse
        plane, $\Delta \phi_{1,2} < 140^{\circ}$;
  \item there must be missing ``longitudinal momenta'' such that
        $ 2 E_e^0 - \sum \left(E - P_z\right) > 12 \GeV $;
  \item at least two jets must be found by the jet cone algorithm each
        with $E_{T,jet} > 5 \GeV$.
\end{enumerate}
Cut (1) suppresses the contamination of $b \bar b$ photoproduction
where the heavy $b$ quarks can undergo semi-leptonic decays.
The transverse energy requirement in cut (2) is high enough to
eliminate contamination from J/$\psi$ photoproduction.
Asking for two isolated $e^{\pm}$ strongly suppresses the main
DIS NC and photoproduction background, leaving only events where either
a photon or a hadronic jet is misidentified as an $e^{\pm}$.
Such a misidentified jet in DIS NC events is most probably
found to be collinear with the true electron in the transverse plane.
%
\begin{figure}[htb]
  \begin{center}
    \begin{tabular}{cc}
      \mbox{\epsfxsize=0.50\textwidth
                             \epsffile{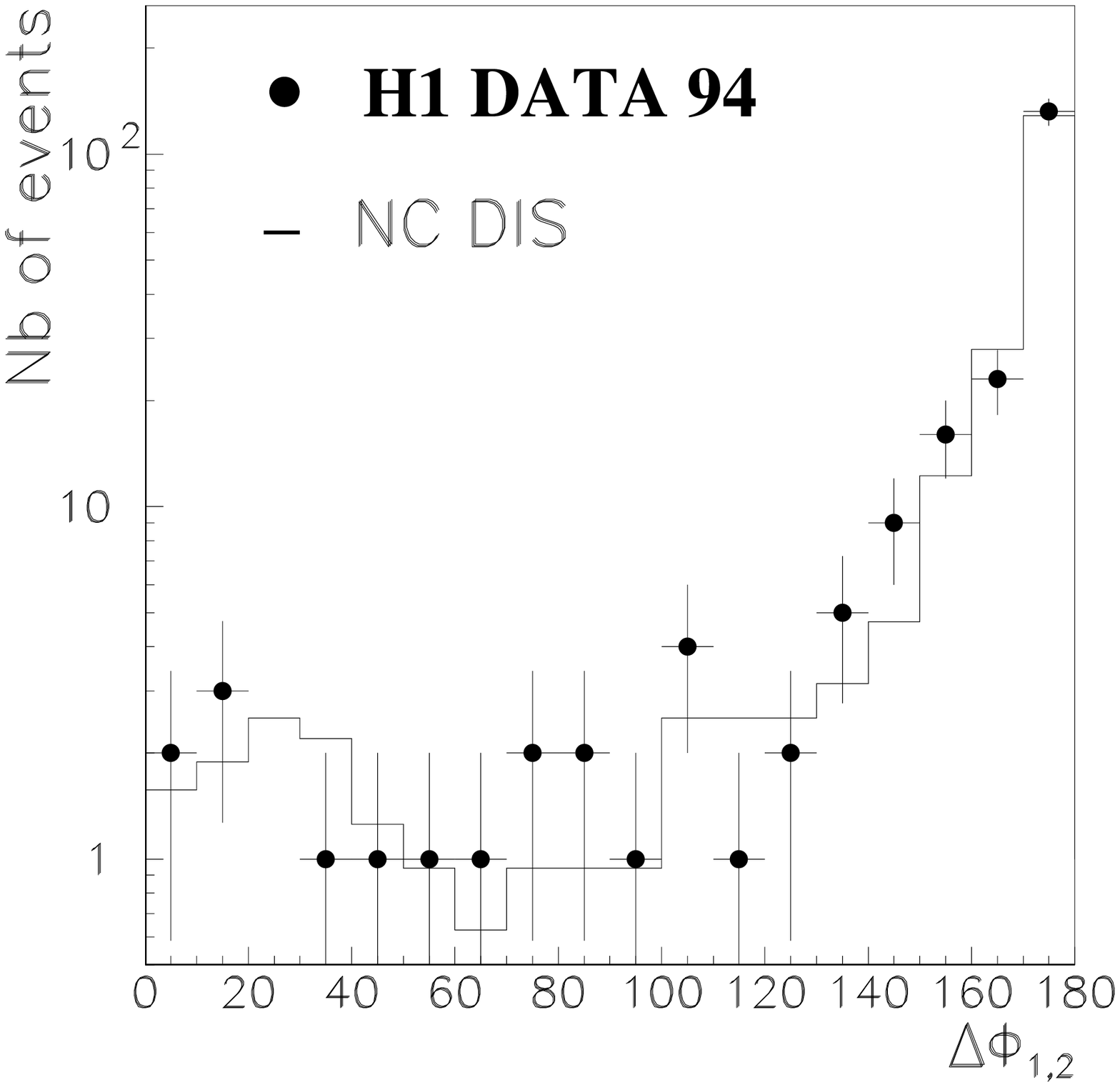}}
      \mbox{\epsfxsize=0.50\textwidth
                             \epsffile{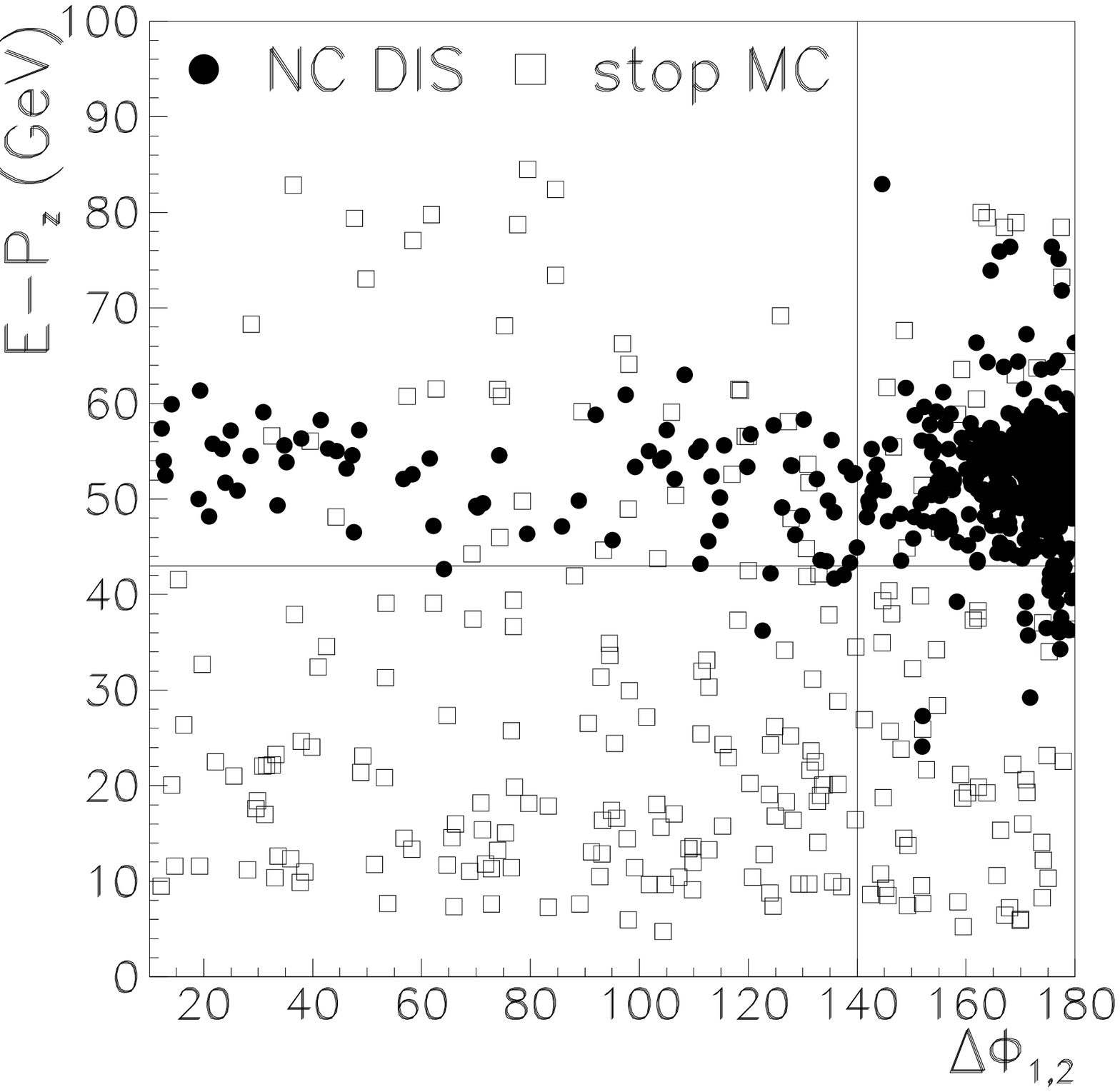}} &
    \end{tabular}
  \end{center}
\caption[]{ \label{fig:stopdis}
 {\small (a) Azimuthal balance $\Delta \phi_{1,2} $ of the two
             $e^{\pm}$ candidates for data (closed dots) and DIS Neutral
             Current Monte Carlo (histogram) when
             $\sum \left(E - P_z\right) > 43 \GeV$.
         (b) Comparison of Monte Carlo expectations for the missing
             longitudinal momentum $\sum \left(E - P_z\right)$
             versus the balance in azimuth of the two $e^{\pm}$
             candidates $\Delta \phi_{1,2} $ for DIS Neutral
             Current (closed dots) and stop pair production in
             $\gamma$-gluon fusion (open squares) for
             $M_{\tilde{t}_1} = 20 \GeV$. }}
\end{figure}
This can be seen in Fig.~\ref{fig:stopdis}a where the azimuthal balance
$\Delta \phi_{1,2}$ between the two $e^{\pm}$ candidates found in the
data is compared with SM Monte Carlo simulation.
For this comparison with DIS NC, an additional cut of
$ \sum \left(E - P_z\right) > 43 \GeV $ is imposed.
The observed acollinearity in $\phi$ between the two $e^{\pm}$
candidates is seen to be well described by the Monte Carlo.
 
The remaining background is suppressed with cut (3) and we are left with
$26.1 \pm 3.9$ expected events from DIS NC while the photoproduction
contamination is negligible.
The estimated DIS NC background agrees well with the $28$ observed
events.
This background is further suppressed by cut (4).
The effect of this cut is seen in Fig.~\ref{fig:stopdis}b showing
Monte Carlo expectations for the correlation between
$ \sum \left(E - P_z\right) $ and the azimuthal balance
$\Delta \phi_{1,2}$ for DIS NC
(plotted for $3 \times {\cal{L}}_{data}$) and stop pair production
(arbitrary normalization).
Two events are rejected by cut (5).
 
With cuts (1) to (5), we are left with no observed event candidate while
$1.0 \pm 0.8$ event is expected from misidentified DIS NC background.
The detection efficiencies for the signal of $\tilde{t}_1$ pair
production after all selection cuts are found to rise from
about $3\%$ at $9 \GeV$ to $32 \%$ at $24 \GeV$.
 
\section{Results }
\label{sec:results}
 
In the absence of significant deviation from the SM expectations, we
now derive exclusion limits for the Yukawa couplings $\lambda'_{1jk}$
as a function of mass, combining all contributing channels and making
use of the number of observed events, of expected background events,
and the signal detection efficiencies for each contributing channel.
 
\hfill \\
\noindent
{\bf Single production of squarks:}
 
For the event topologies {\large S1} to {\large S3}, the detection
efficiencies are folded with a mass bin of variable width which slides
over the accessible $M_{\tilde q}$ range.
The bin width is optimized taking into account the mean expected
background and the expected $\tilde{q}$ mass resolution and contains
about $68\%$ of the signal at a given mass,
e.g. at $150 \GeV$ we typically have a full bin width of
$\Delta M_e \simeq 25 \GeV$ in {\large S1}, of
$\Delta M_h \simeq 35 \GeV$ in {\large S2}, and of
$\Delta M_{dec} \simeq 40 \GeV$ in {\large S3}.
For event topologies {\large S4} to {\large S8} where the number of
observed events and expected background is always $< 10$, the signal
is integrated above the selection cuts (i.e. without explicit
restriction on the reconstructed mass).
More details on the methodology for the limits derivation and on the
procedure for folding the channel per channel statistical and
systematic errors are given in~\cite{H1LQ94}.
 
The detection efficiencies have to be folded with the branching fractions
in each of the possible event topologies, properly taking into account
the relative production cross-section of the various squark flavours
(i.e. $\bar{\tilde{d}_R}$ and $\tilde{u}_L$ for $\lambda'_{111} \ne 0$).
For small $M_{\chi_1^0}/M_{\tilde{q}}$ and/or small $\lambda'_{1jk}$
values where gauge decays of the squark are expected to dominate,
the dependence upon the values of the free parameters of the MSSM
has to be fully considered for the coupling constants at the
$\tilde{q} \rightarrow q + \chi_1^0$ or
$\tilde{q} \rightarrow q' + \chi_1^+$ vertex as well as the decay
branchings of the $\chi_1^0$ and $\chi_1^+$.
We will discuss our results in terms of the usual parameters:
the ratio $\tan \beta$ of the two Higgs field vacuum expectation
values, the higgsino mixing parameter $\mu$, the mass parameter $M_2$
for the $SU(2)$ gauginos.
 
\begin{figure}[htb]
  \begin{center}
    \begin{tabular}{cc}
      \mbox{\epsfxsize=8.0cm \epsffile{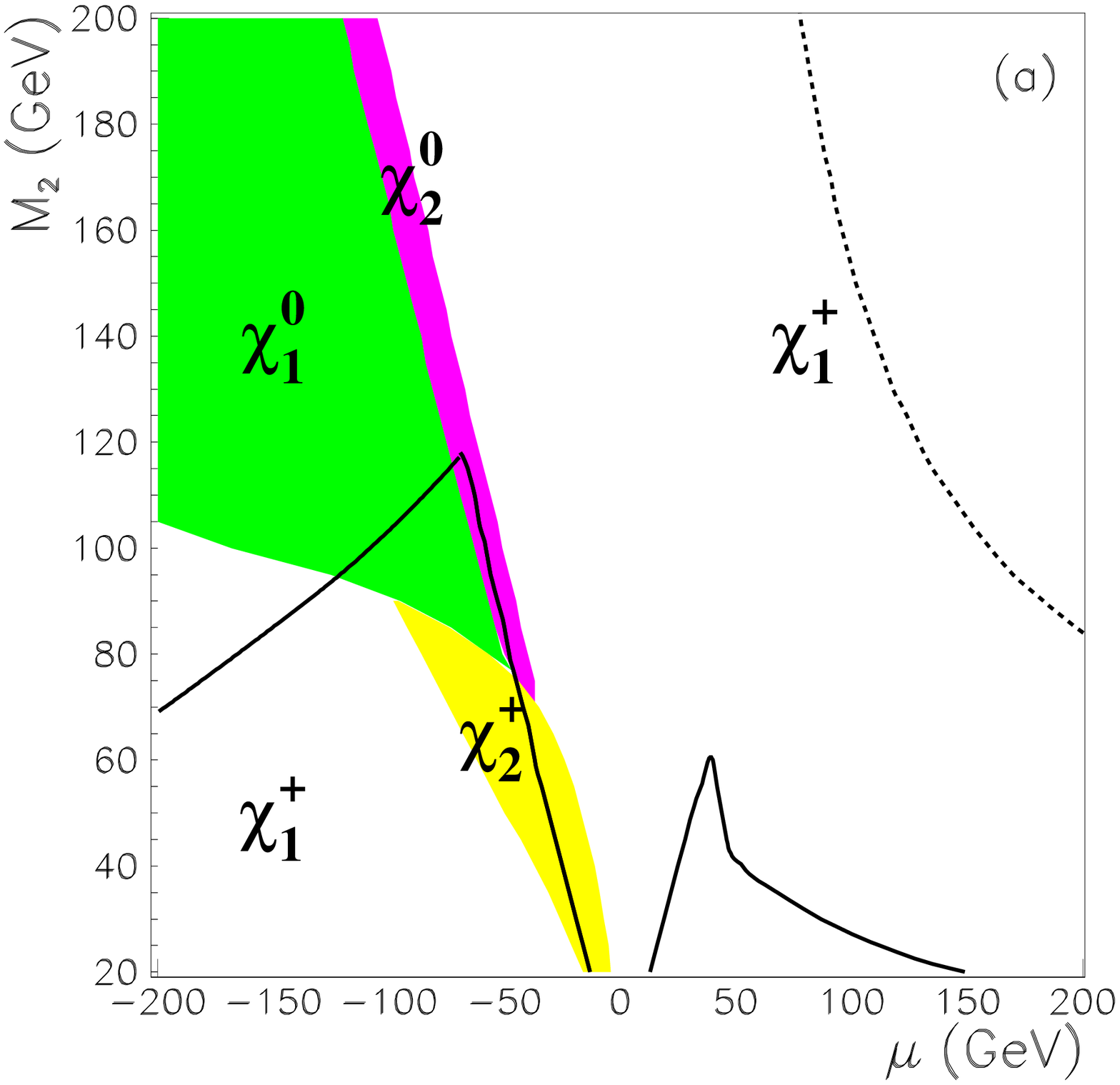}}
      \mbox{\epsfxsize=8.0cm \epsffile{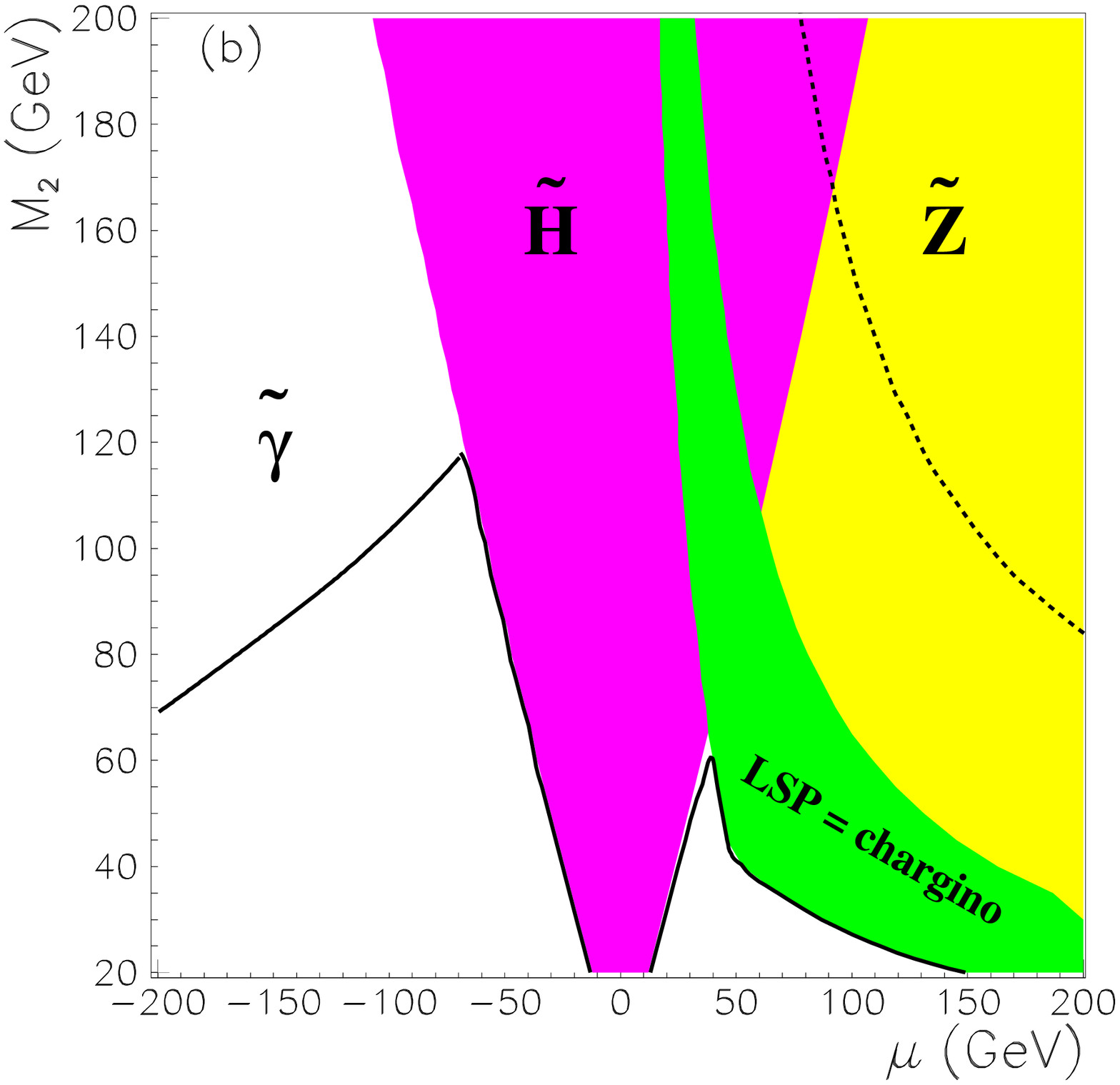}}
    \end{tabular}
  \end{center}
 \caption[]{ \label{fig:lspul}
   {\small  (a) Regions in the $M_2,\mu$ plane where a
            $150 \GeV$ $\tilde{u_{L}}$ squark decays dominantly
            into one of the $\chi_i^0$ or $\chi_i^+$ states
            for $\tan \beta = 1$;
            (b) Main component of the lightest neutralino in
            the plane ($M_{2},\mu$), for $\tan \beta = 1$.
            In (a) and (b) the region below the full curve corresponds 
            to a domain in which the branching ratio of the $\chi_1^0$ 
            into $e^{\pm} q \bar{q^{\prime}}$ is greater than $80\%$.
            Along the dotted curve, the branching into
            $e^{\pm} q \bar{q^{\prime}}$ is of about $20\%$ for a
            $\chi_1^0$ of $20 \GeV$. }}
\end{figure}

The domains of the ($M_2,\mu$) plane where decays into a specific
neutralino or chargino dominate for the $\tilde{u}_L$ are shown
in~Fig.\ref{fig:lspul}a for a squark mass in the middle of the
accessible $M_{\tilde{q}}$ range.
%
The decay into the lightest chargino $\chi_1^+$ is seen to dominate
as soon as kinematically allowed.
This is contrary to the $\bar{\tilde{d}_R}$ which mainly decays into
$\bar{d} + \chi_1^0$ (except when the $\chi_1^0$ is
$\tilde{H}$-dominant).
For a $\tilde{H}$-dominant LSP the coupling constant is
small enough to suppress strongly squark gauge decays into $\chi_1^0$.
 
Regions for various kinds of $\chi_1^0$ in the parameter space
are shown in Fig.~\ref{fig:lspul}b.
Similar plots of the admixture of the weak eigenstates
$\tilde{\gamma}, \tilde{Z}, \tilde{H}_1^0$ and $\tilde{H}_2^0$ in the
neutralino LSP are given in~\cite{DREINERM}.
In the region where the $\chi_1^0$ is $\tilde{\gamma}$-like,
the branching ratio of the $\chi_1^0$ into $ e^{\pm} q \bar{q'} $
is greater than $\approx 60\%$ for $\tan \beta=1$.
When the $\chi_1^0$ is a pure $\tilde{\gamma}$ ($M_2 = 0$),
this branching saturates at about $88\%$.
In the region where the $\chi_1^0$ is $\tilde{Z}$-like,
for instance along the dotted line for a $\chi_1^0$ of $20 \GeV$,
the branching ratio into $ e^{\pm} q \bar{q'} $ is about $20\%$
 
%
%
\begin{table}[htb]
 \renewcommand{\doublerulesep}{0.4pt}
 \renewcommand{\arraystretch}{1.2}
 \begin{center}
  \begin{tabular}{||c|c|c|c|c|c|c|c|c|c|c||}
  \hline \hline
   $M_{\tilde{q}}$ ; $M_{\chi_1^0}$ &
   & $S_{1}$  & $S_{2}$ & $S_{3}$ & $S_{4}$ & $S_{5}$
   & $S_{6}$ & $S_{7}$ & $S_{8}$ & $\sum \varepsilon {\cal B}$   \\
    ($\GeV$) &  &  &  &  &  &  &  &  &  &             \\
    \hline  \hline
 ${\mbox{ }}$  & $N_{observed}$
  & 20    &17     &60    &0     &9    &3     &0     &4 &$ $\\
 $ 75 \; ; \; 40 $ & $N_{expected}$
  & 20.66 & 22.9  & 41.4 & 0.  & 3.9 & 4.07  & 0.76  & 2.4 &$ $\\
          &${\cal B}$($\tilde{\gamma}$)
 & 1.2    &0.4    &40.8  &40.8  &16.9  &-   &0    &0  &25.5 \\
 $\mbox{ }$ &${\cal B}$($\tilde{Z}$)
 & 11.2    &5.0    &5.4   &3.0   &71.4  &-  &0.2   &1.7   &13.3 \\
 $\mbox{ }$ &${\cal B}$($\tilde{H}$)
 & 89.9    &10.0    &0   &-   &0  &0  &-   &-   &37.0 \\
 $\mbox{ }$ & $\varepsilon(\tilde{\gamma})$
 & 35.7   & 48.8  &34.8  &24.3  &4.7   &0  &0    &0   &$ $\\
 \hline \hline
%
%
 ${\mbox{ }}$  & $N_{observed}$
 & 3     &8      &5     &0     &9    &3     &0     &4 &$ $\\
 $ 150 \; ; \; 40 $ & $N_{expected}$
 & 1.62   &8.58   &2.71  &0.   &3.9   &4.07    &0.76   &2.4 &$ $\\
            & ${\cal B}$($\tilde{\gamma}$)
 & 3.4    &1.0    &26.6  &7.2  &17.0  &-   &7.5    &1.5  &20.1 \\
 $\mbox{ }$ & ${\cal B}$($\tilde{Z}$)
 & 7.3    &1.8    &16.7   &0.5   &45.0  &-  &0.7    &5.8   &28.2 \\
 $\mbox{ }$ & ${\cal B}$($\tilde{H}$)
 & 10.0    &1.2    &5.4   &-     &23.2  &7.8  &-   &-   &21.0 \\
 $\mbox{ }$ & $\varepsilon(\tilde{\gamma})$
 & 45.5   & 54.2  &33.1  &20.9   &32.8   &$51.0^*$  &23.1    &29.8   &$ $\\
 \hline \hline
%
%
 ${\mbox{ }}$  & $N_{observed}$
 & 0     &1      &17     &0     &9    &3     &0     &4 &$ $\\
 $ 250 \; ; \; 40 $ & $N_{expected}$
 & 0.0    &0.59   &16.2  &0.   &3.9   &4.07    &0.76   &2.4 &$ $\\
           & ${\cal B}$($\tilde{\gamma}$)
 & 36.5    &4.9    &20.4  &2.5  &18.7  &-   &0.3    &0.06  &30.8 \\
 $\mbox{ }$ & ${\cal B}$($\tilde{Z}$)
 & 43.3    &4.2    &15.9   &0.2   &19.7  &-  &0.04    &0.3   &32.6 \\
 $\mbox{ }$ & ${\cal B}$($\tilde{H}$)
 & 45.6    &4.3    &4.2   &-   &0.6  &0.2  &-   &-   &23.8 \\
 $\mbox{ }$ & $\varepsilon(\tilde{\gamma})$
 & 46.1   & 29.9  &25.5  &14.3   &36.7   &$43.1^*$  &24.0    &30.5  &$ $ \\
 \hline \hline
%
%
 ${\mbox{ }}$  & $N_{observed}$
 & 3     &8      &6     &0     &9    &3     &0     &4 &$ $\\
 $ 150 \; ; \; 80 $ & $N_{expected}$
 & 1.62   &8.58   &3.48  &0.   &3.9   &4.07    &0.76   &2.4 &$ $\\
          & ${\cal B}$($\tilde{\gamma}$)
 & 23.1    &1.2    &26.3  &26.3  &23.0  &-   &0    &0  &41.4 \\
 &${\cal B}$($\tilde{Z}$)
 & 80.0    &1.6    &1.9   &1.9   &14.5  &-  &0    &0   &46.1 \\
 $\mbox{ }$ &${\cal B}$($\tilde{H}$)
 &21.4     &1.4    &7.4   &0   &13.3  &4.4  &-   &-   &17.6 \\
 $\mbox{ }$ & $\varepsilon(\tilde{\gamma})$
 & 45.5   & 54.2  &42.9  &27.0   &51.6   &$36.5^*$  &0    &0   &$ $\\
 \hline \hline
%
%
 ${\mbox{ }}$  & $N_{observed}$
 & 0     &1      &6     &0     &9    &3     &0     &4 &$ $\\
 $ 250 \; ; \; 80 $ & $N_{expected}$
 & 0.     &0.59   &3.10  &0.   &3.9   &4.07    &0.76   &2.4 &$ $\\
           & ${\cal B}$($\tilde{\gamma}$)
 & 52.1    &4.4    &2.5  &1.8  &2.2  &-   &0.3    &0.13  &28.1 \\
 &${\cal B}$($\tilde{Z}$)
 & 56.5    &4.3    &9.6   &0.06   &14.1  &-  &0.04    &1.41   &40.0 \\
 $\mbox{ }$ & ${\cal B}$($\tilde{H}$)
 &58.2     &4.6    &3.5   &0   &0.2  &0.08  &-   &-   &29.6 \\
 $\mbox{ }$ & $\varepsilon(\tilde{\gamma})$
 & 46.1   & 29.9  &36.3  &20.3   &58.6   &$33.0^*$  &44.5    &44.7 &  \\
\hline \hline
\end{tabular}
   \caption[]
          {\small \label{tab:sqresu}
              Number of observed events, of expected background
              events and branching ratios ${\cal B}$ (in $\%$) 
              corresponding
              to channels {\large{S1}} to {\large{S8}} for some values
              of $M_{\tilde{q}}$ and $M_{\chi_1^0}$, and presented
              for $\tilde{\gamma}$-like, $\tilde{Z}$-like and
              $\tilde{H}$-like $\chi_1^0$.
              The efficiencies $\varepsilon$ are given in each
              channel for the $\tilde{\gamma}$-like case except$^*$
              for {\large{S6}} where they are given for the
              $\tilde{H}$-like case.
              Also given is the $\sum \varepsilon {\cal B}$ summed over
              all channels for each $\chi_1^0$ type. }
 \end{center}
 
\end{table}
\vfill \clearpage
The relative contribution of each of the channels {\large{S1}}
to {\large{S8}} is given in Table~\ref{tab:sqresu} for a few
representative cases.
For the chosen examples, the mass of the $\chi^+_1$ is about twice
that of the $\chi_1^0$ expect for heavy (e.g. $80 \GeV$)
$\tilde{H}$-like $\chi_1^0$ for which
$M_{\chi^+_1}/M_{\chi_1^0} \sim 1.3$.
In the low squark mass region, the decay modes
{\large{S3}} and {\large{S4}} dominate since the decay of the squark
into $\chi_1^+$ is largely suppressed by phase space and since
the Yukawa couplings probed are small.
For medium masses, when the $\tilde{q}$ is heavier than
the $\chi^+_1$, it decays dominantly via {\large{S3}} and {\large{S5}}.
The channel {\large{S5}} contributes mainly when the $\chi_1^0$ is
$\tilde{Z}$-like, since in that case the branching
ratio for $\chi_1^0 \rightarrow \nu + 2jets$ is generally above $70\%$.
At very large $\tilde{q}$ masses, we are only sensitive to high values
of the coupling, so that the \Rp\ decay {\large{S1}} into
$e^+ \, + \, jet$ dominates whereas \Rp\ decays {\large{S2}} are
strongly suppressed by the parton density.
 
From the total branching fraction (i.e. summed over all above channels),
it is inferred that the contribution of the decays into heavier
$\chi_i^0$($i > 1$) and $\chi_j^+$ ($j>1$) is generally small.
Hence, in order to simplify the derivation of limits we have assumed
conservatively that the $\tilde{q}$ decay into these heavy states
are allowed but measured with vanishingly small efficiencies.
Folded in the derivation of limits are systematic errors coming from
the uncertainty on the luminosity measurement (1.5\%),
the finite Monte Carlo statistics,
the absolute energy calibration which leads to an uncertainty on
the background estimation of about 10\%,
the choice of the scale entering the structure function calculation
(which leads to $\approx 7\%$ uncertainties in the cross-section) and
the choice of the parton density parametrization.
 
\begin{figure}[htb]
 \begin{center}
    \mbox{\epsfxsize=0.8\textwidth \epsffile{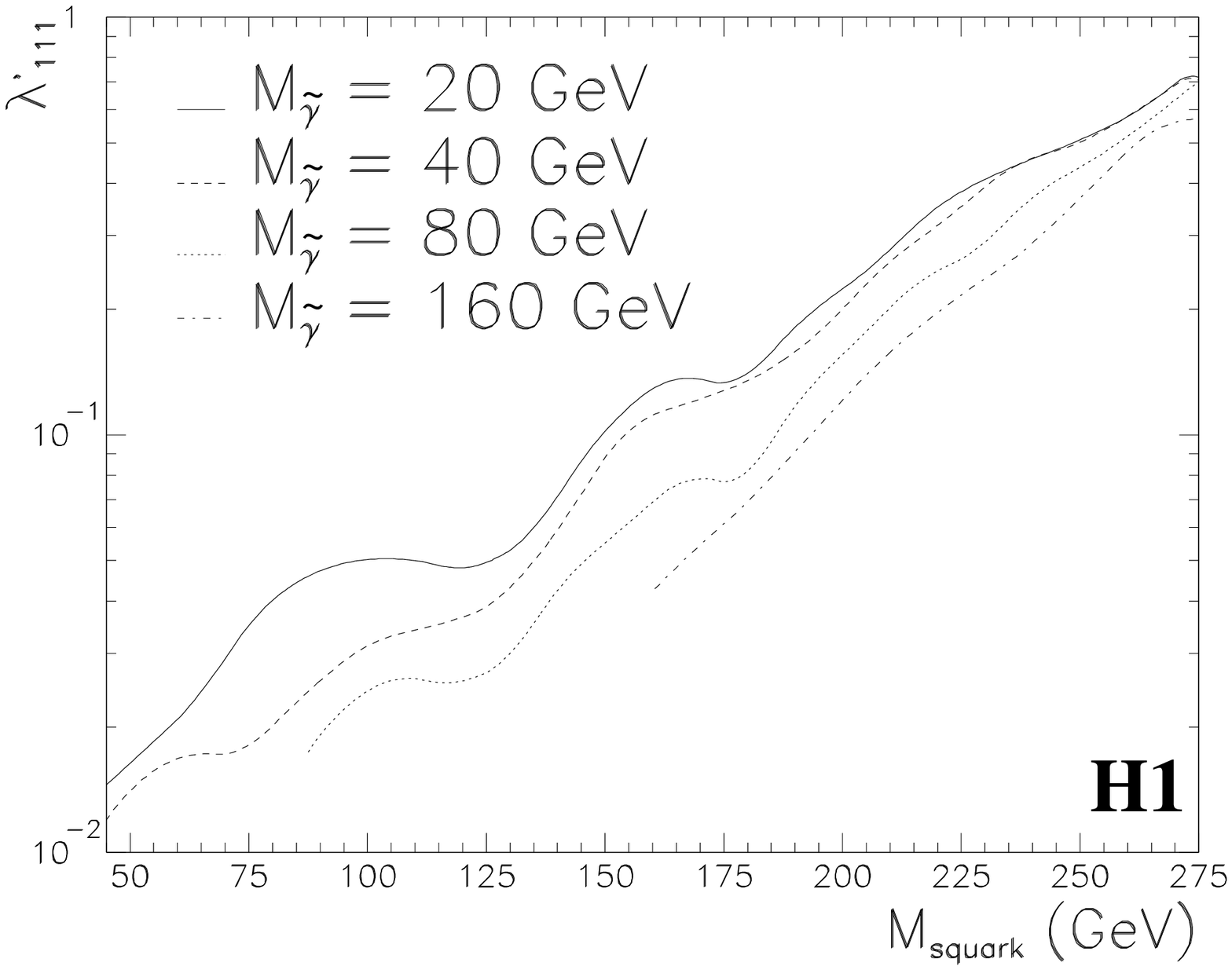 }}
 \end{center}
\caption[]{ \label{fig:sulim94}
 {\small    Exclusion upper limits at 95\% CL for the coupling
            $\lambda'_{111}$ as a function of the squark
            mass for various fixed photino masses and derived for
            $\tan \beta = 1$. The regions above
            the curves are excluded.
            The limits combine \Rp\ and gauge decays of
            the $\bar{\tilde{d}}_R$ and
            $\tilde{u}_L$. 			}}
\end{figure}

%
The exclusion limits obtained on the coupling $\lambda'_{111}$ at
$95\%$ confidence level (CL) are shown in Fig.~\ref{fig:sulim94} as
function of the $\tilde{q}$ mass ($\tilde{u}_L$ and
$\bar{\tilde{d}}_R$) in the hypothesis that the $\chi_1^0$ is a pure
$\tilde{\gamma}$.
The limits represent an improvement of about a factor two to three
compared to our previously published results~\cite{H1LQ94} at low squark
masses. The gain from integrated luminosity is only partially
cancelled by a less favourable quark density since for incident positron,
the $e^+ d \rightarrow \tilde{u}_L$ production dominates
over $e^+ \bar{u} \rightarrow \bar{\tilde{d}}_R$ whilst
$e^- u \rightarrow \tilde{d}_R$ dominates over
$e^- \bar{d} \rightarrow \bar{\tilde{u}}_L $
for incident electron data.
The limits also improve at largest mass where the smaller coupling
probed implies a narrow observable resonance width.
 
For the smallest couplings accessible here, a squark lighter than
$\approx 230 \GeV$ undergoes dominantly a gauge decay.
Hence in contrast to earlier searches~\cite{H1LQ94}, we are here
sensitive to event topologies immediately distinguishable from those
of leptoquarks.
The existence of first generation squarks with \Rp\ Yukawa
coupling $\lambda'_{111}$ is excluded for masses up to $ 240 \GeV $
(depending on the $\chi_1^0$ mass) at coupling strengths
$\lambda'^2_{111} / 4\pi \, \gtrsim \alpha_{em} $
( up to $ 130 \GeV $ for coupling strengths
$\gtrsim 0.01 \times \alpha_{em} $).
 
\begin{figure}[htb]
 \begin{center}
     \mbox{\epsfxsize=0.8\textwidth \epsffile{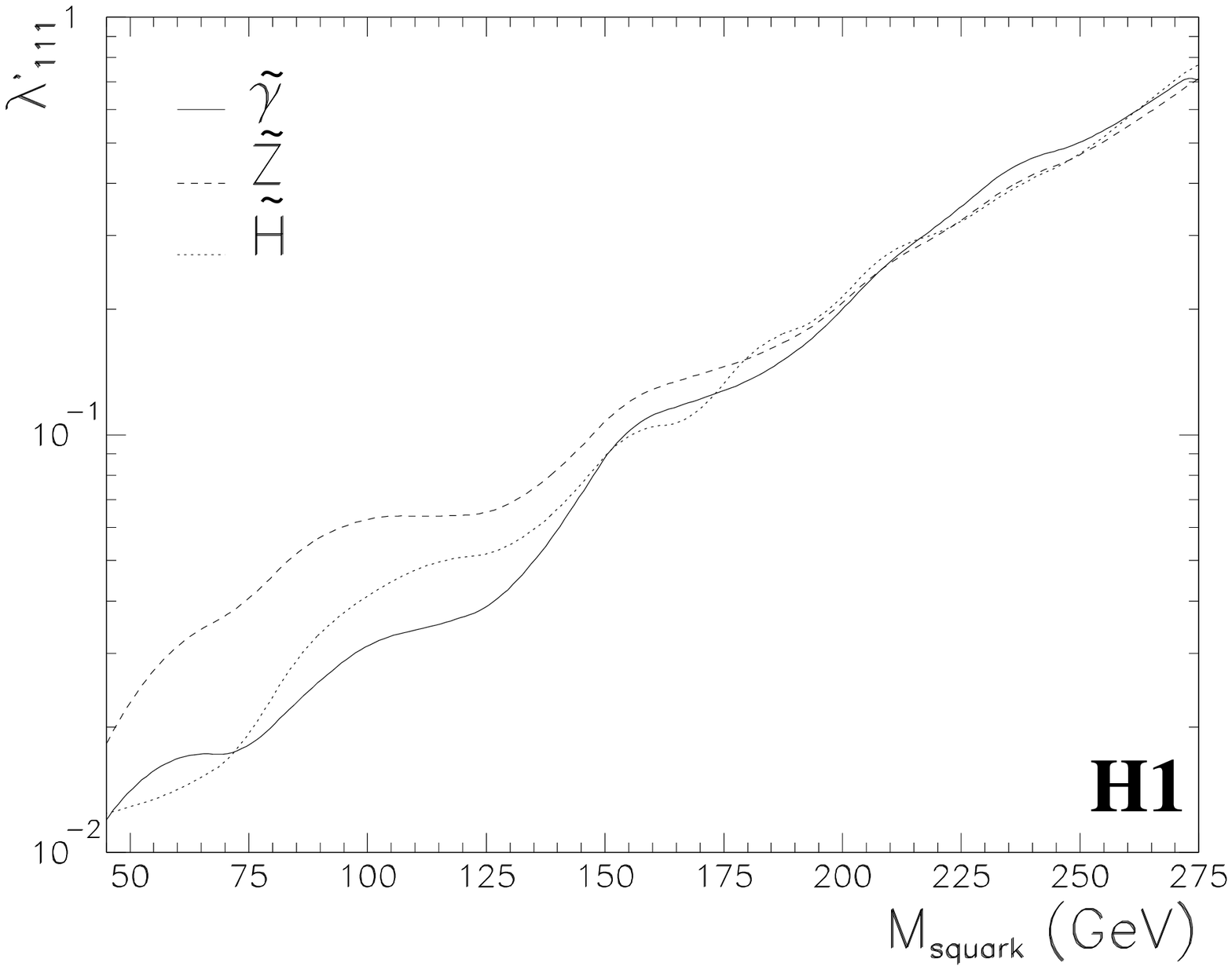}}
 \end{center}
\caption[]{ \label{fig:lspnat}
 {\small
     Exclusion upper limits at 95\% CL for the coupling
     $\lambda'_{111}$ as a function of squark mass for
     different natures of the $\chi_1^0$ for $M_{\chi_1^0} = 40 \GeV$
     and for $\tan \beta = 1$ (region above the curve excluded).
     The full curve corresponds to cases where the LSP is
     $\tilde{\gamma}$-like, the dotted one to a
     $\tilde{H}$-dominant
     LSP
     and the dashed one to a $\tilde{Z}$-like LSP.}}
\end{figure}

In the more general case, where the $\chi_1^0$ is a mixture of
gauginos and higgsinos, the exclusion limits on the coupling
$\lambda'_{111}$ at $95\%$ CL
are shown in Fig.~\ref{fig:lspnat} as function of the $\tilde{q}$ mass.
%
Here, the limits are derived for a reference point in the MSSM
parameter space chosen as $\mu = -160 \GeV$ and $M_2 = 60 \GeV$ for a
$\tilde{\gamma}$-like $\chi_1^0$, $\mu = 150 \GeV $ and $M_2 = 150 \GeV$
for a $\tilde{Z}$-like $\chi_1^0$, and
$\mu = -44 \GeV $ and $M_2 = 140 \GeV$ for a
$\tilde{H}$-like $\chi_1^0$.
These points in the parameter space lie outside the domain excluded
from the invisible $Z_0$ width measurement at LEP~\cite{ALEPHRP}.
The rejection limits for a $\tilde{\gamma}$-like $\chi_1^0$ are found
not to differ much from those obtained for pure $\tilde{\gamma}$.
The limits are also seen not to depend too strongly on the nature
of the LSP.
The three curves merge together at highest squark masses, where the
branching of the squark into $\chi_1^0 q$ becomes negligible
relative to \Rp\ decay.
The limits obtained are found moreover not to depend strongly on the
parameter $\tan \beta$.
Varying $\tan \beta$ from $1$ to $40$, we find that the limits
only slightly degrade and mainly at very low $\tilde{q}$ masses,
by $30\%$ at $M_{\tilde{q}} = 45 \GeV$ down to $2\%$ at $200 \GeV$.
 
From the analysis of the $\lambda'_{111}$ case involving the
$\bar{\tilde{d}}_R$ and  $\tilde{u}_L$ squarks, limits can be
deduced on the $\lambda'_{1jk}$ by folding in the proper parton
densities. Such limits are given in Table~\ref{tab:limlam} at
$M_{\tilde{q}}=150 \GeV$.
For $M_{\tilde{q}} \gtrsim 150 \GeV$, the exclusion limits for 
$\lambda'_{111}$ and $\lambda'_{121}$ in particular are found to 
coincide within $5\%$.
\begin{table*}[hbt]
 \begin{center}
 \begin{tabular}{p{0.40\textwidth}p{0.60\textwidth}}
   \vspace{-3.0cm}
   \caption[] {\label{tab:limlam}
         {\small
         Exclusion upper limits at $95\%$ CL on the couplings
         $\lambda'_{1jk}$ for $M_{\tilde q} = 150 \GeV$ and
         $M_{\chi_1^0} = 80 \GeV$. The quoted values for
         $\lambda'_{111}$ are given for $\tilde{\gamma}$-dominant
         and for $\tilde{Z}$-dominant $\chi_1^0$.
         In other cases, the higgsino component of the $\chi_1^0$
         is assumed to be vanishingly small.
         Moreover, the results for cases with $j = 3$ are only
         valid under the additional restriction that
         $M_{\chi_1^+} > M_{\tilde q}$. }} &
   \begin{tabular}{|c|c|c|}                 \hline
     $\lambda'_{1jk}$
        & $\lambda'_{lim} \mbox{     } \tilde{\gamma}$ case
        & $\lambda'_{lim} \mbox{     } \tilde{Z}$ case
                                                       \\ \hline
     $\lambda'_{111}$    & 0.056         & 0.048    \\ \hline
     $\lambda'_{112}$    & 0.14          & 0.12     \\ \hline
     $\lambda'_{113}$    & 0.18          & 0.15     \\ \hline
     $\lambda'_{121}$    & 0.058         & 0.048    \\ \hline
     $\lambda'_{122}$    & 0.19          & 0.16     \\ \hline
     $\lambda'_{123}$    & 0.30          & 0.26     \\ \hline
     $\lambda'_{131}$    & 0.06          & 0.05     \\ \hline
     $\lambda'_{132}$    & 0.22          & 0.19     \\ \hline
     $\lambda'_{133}$    & 0.55          & 0.48     \\ \hline
   \end{tabular}
 \end{tabular}
\end{center}
\end{table*}
 
Our rejection limits extend considerably beyond the only other
collider limits of $M_{\tilde{q}}>100 \GeV$ for $\lambda'_{111}$
inferred in~\cite{ROY} from dilepton data of the Tevatron experiments.
Moreover, this Tevatron limit was derived only under the restrictive
assumption that the LSP is the $\tilde{\gamma}$ and that squarks
other than the stop are degenerate in mass.
If one assumes that one squark is substantially lighter than
the others, this bound is much weaker~\cite{RPVIOLATION} and
only slightly above the mass reach of LEP 1.
On the contrary, our $\lambda'_{111}$ coupling limits only weakly
depend on the mass degeneracy assumption since, as mentioned above,
the $\tilde{u}$ production strongly dominates.
Hence, if the first generation $\tilde{u}$ is significantly
lighter than other squarks, we probe here (and similarly for
the $\tilde{d}$ in~\cite{H1LQ94}) a large portion of the
mass-coupling domain unexplored by other experiments.
 
There are no other direct limits published for $\lambda'_{1jk}$
($j$ or $k \ne 1$).
This is particularly interesting since indirect limits for this
coupling are also weaker~\cite{RPVIOLATION} than 
those~\cite{HIRSCH} for $\lambda'_{111}$.
 
\hfill \\
\noindent
{\bf Production of stop squarks:}

For the derivation of exclusion limits for the pair production of
stop squarks, a $15\%$ uncertainty on the cross-section (which varies
from $\sigma_{stop} \sim 200 \picob$ at $9 \GeV$
to   $\sigma_{stop} = 1 \picob$ at $24 \GeV$) is due to the
specific choice of gluon density.
The uncertainty was determined by comparing with MRSD--~\cite{MRSDM} and
constitutes the main source of systematic error.
 
\begin{figure}[htb]
  \begin{center}
     \mbox{\epsfxsize=0.9\textwidth \epsffile{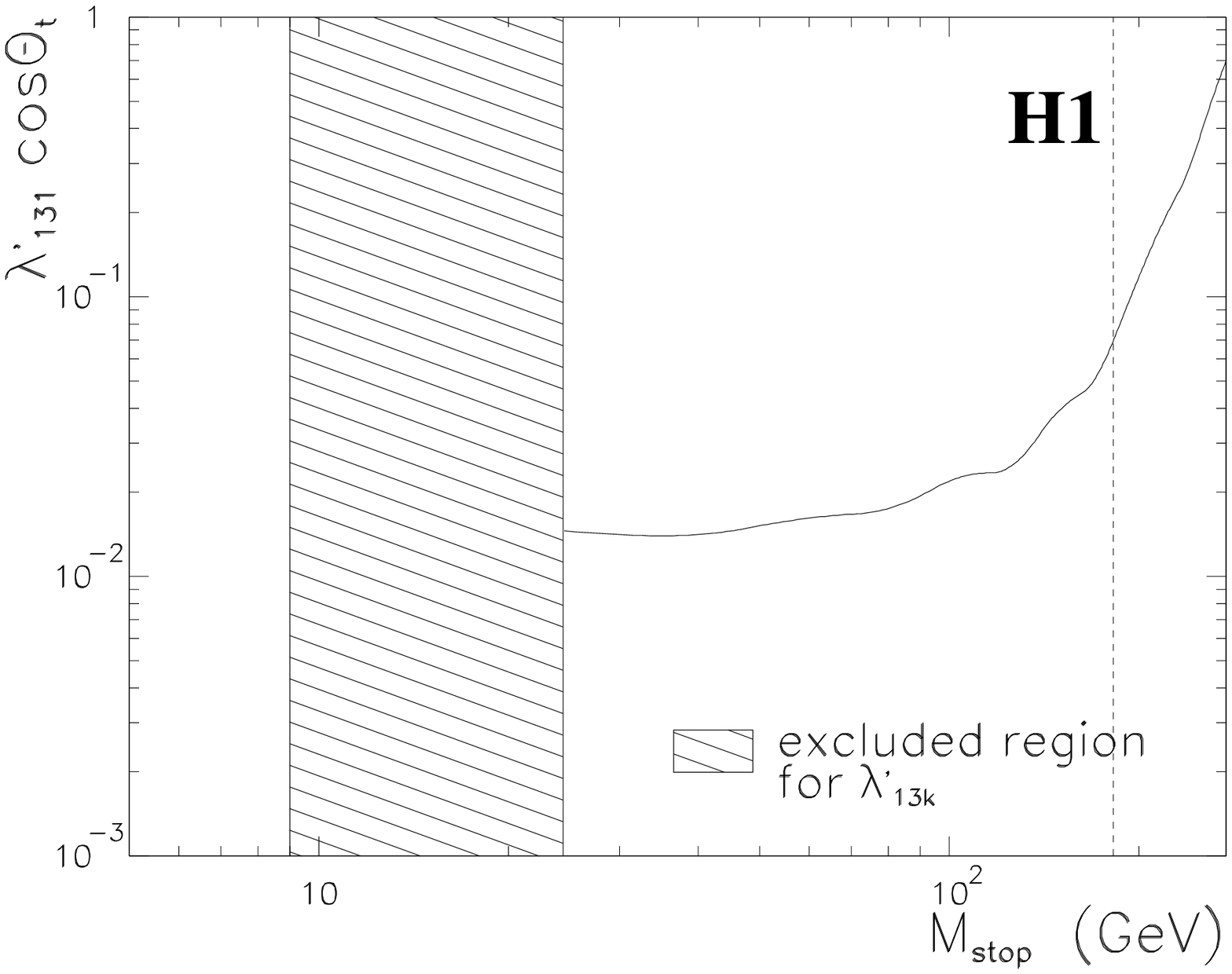}}
  \end{center}
\caption[]{ \label{fig:stopair}
 {\small    Exclusion limits for the coupling
            $\lambda'_{131} \times \cos \theta_t $
            as function of the stop mass.
            The exclusion curve is obtained from the single stop
            production (region above the curve excluded).
            Beyond the dashed line, the limits are only
            valid for  $M_{\tilde{t}} < M_{top} + M_{\chi_1^0}$.
            The hatched domain is excluded by the pair production
            search and concerns $\lambda'_{13k}$ with $k=1,2$. }}
\end{figure}

The exclusion limit obtained on the stop mass at
$95\%$ CL is shown in Fig.~\ref{fig:stopair}.
%
A stop in the range $ 9 < M_{\tilde{t}_1} < 24.4 \GeV $
is excluded at $95\%$ CL.
This limit does not depend upon the value of $\lambda'_{13k}$ (as far
as $\lambda'_{13k} \times \cos \theta_t \gtrsim 10^{-4}$, below which
value the decay into $c$ and $\chi_1^0$ cannot be neglected anymore).
The angle $\theta_t$ is the mixing angle in the mass matrix of the
stop (see for example~\cite{KON}).
 
If the coupling $\lambda'_{131}$ dominates, the stop can
be singly produced in reactions of the type
$e^+ + d \rightarrow \tilde{t}$. Under our phenomenological
assumptions (see section~\ref{sec:pheno}) and as long as
$(\lambda'_{131} \times \cos \theta_t) \gtrsim 10^{-4}$,
the search of the $\tilde{t}$ borrows from the
analysis for \Rp\ decays of first generation squarks.
We find that masses below $138 \GeV$ are excluded at
$95 \%$ CL for coupling strength of
$(\lambda'_{131} \times \cos \theta_t)^2/4\pi \, \gtrsim
                                                   0.01 \alpha_{em} $.
This represents an increase of sensitivity of about an order of
magnitude compared to our previous results~\cite{H1LQ94}.
As a comparison with other experiments, a $\tilde{t}$ lighter than
$38 \GeV$ is excluded at $95 \%$ CL from LEP data~\cite{OPAL} for
$\theta_{t} = 0$ from the width of the $Z_0$.
But for $\theta_{t}$ close to the value for which the $Z_0$ decouples
from the stop ($ 0.7 < \theta_{t} < 1.4$) there are no existing
limit from LEP for \Rp\ stops.
 
The coupling limit in Fig.~\ref{fig:stopair} is extended beyond
$M_{\tilde{t}} \simeq M_{top}$. This portion of the exclusion limit
curve is only valid for  $M_{\tilde{t}} < M_{top} + M_{\chi_1^0}$.
 
%
\section{Conclusions}
 
We have searched for squarks from $R$-parity violating supersymmetry.
The search was carried out for the first time at HERA in all possible
decay processes allowed when wandering in the parameter space of the
Minimal Supersymmetric Model.
No significant evidence for the production of squarks was
found and mass dependent limits on the couplings were derived.
The existence of first generation squarks at masses up to
$240 \GeV$ are excluded at $95\%$ confidence level for a strength of
the Yukawa coupling
$\lambda'_{111}$ of $\lambda'^2_{111} / 4\pi \, = \alpha_{em} $.
The limits extend far beyond results obtained at other colliders
where our excluded domain in the mass-coupling plane for masses
$\gtrsim 100 \GeV$ has never been explored.
 
Scalar stop squarks were searched in pair and single production modes.
The existence of light scalar stops with $\lambda'_{13k}$ couplings
to light fermions is excluded for masses $9 < m_{\tilde{t}} < 24.4 \GeV$
at $95\%$ confidence level.
Stop squarks with $\lambda'_{131}$ couplings are excluded below
$138 \GeV$ at $95 \%$ confidence level for couplings
$(\lambda'_{131} \times \cos \theta_t)^2/4\pi \, \gtrsim
                                                   0.01 \alpha_{em} $.
 
\section*{Acknowledgements}
We wish to thank the HERA machine group as well as the H1 engineers and
technicians who constructed and maintained the detector for their
outstanding efforts.
We thank the funding agencies for their financial support.
We wish to thank the DESY directorate for the support
and hospitality extended to the non-DESY members of the collaboration.
 
 
{\Large\normalsize}


\begin{thebibliography}{99}
 
%
%
 
\bibitem{H1LQ94}
 H1 Collaboration, T.~Ahmed et al., Z.~Phys.~C64 (1994) 545.
\vspace{-2mm}
 
\bibitem{RPVIOLATION}
J.~Butterworth and H.~Dreiner, Nucl. Phys. B397 (1993) 3,
and references therein.
\vspace{-2mm}
 
\bibitem{H1LQ95}
H1 Collaboration, S.~Aid et al., Phys.~Lett.~B369 (1996) 173.
\vspace{-2mm}
 
\bibitem{BUCHMULL}
W.~Buchm\"uller, R.~R\"uckl and D.~Wyler, Phys.~Lett.~B191 (1987) 442.
\vspace{-2mm}
 
\bibitem{SUSY95}
E.~Perez and Y.~Sirois, ``SUSY Searches at HERA'',
To be published in the Proceedings of the
International Workshop on Supersymmetry and Unification
of Fundamental Interactions,
Editions Fronti\`eres (I.~Antoniadis and H.~Videau, Editors),
(May 1995, Palaiseau).
\vspace{-2mm}
 
\bibitem{DRPEREZ}
E.~Perez, ``Recherche de Particules en Supersym\'etrie Violant la
R-parit\'e dans H1 \`a HERA'', Th\`ese de Doctorat,
DAPNIA/SPP report (in French), to be published.
\vspace{-2mm}
 
\bibitem{GUNION}
J.F.~Gunion and H.E.~Haber, Nucl. Phys. B272 (1986) 1.
\vspace{-2mm}
 
\bibitem{KONRP}
T.~Kon, T.~Kobayashi and S.~Kitamura, Phys. Lett. B333 (1994) 263;
T.~Kon et al., Z. Phys. C61 (1994) 239.
 
\vspace{-2mm}
 
 
\bibitem{H1DETECT}
H1 Collaboration, I.~Abt et al., ``The H1 Detector at HERA'',
DESY preprint 93-103 (July 1993) 194pp;
{\it idem} DESY Internal Report H1-96-01 (March 1996) 157pp.
 
\vspace{-2mm}
 
\bibitem{H1LARCAL}
H1 Calorimeter Group, B.~Andrieu et al.,
Nucl. Instr. and Meth. A336 (1993) 460.
\vspace{-2mm}
 
\bibitem{H1CALRES}
H1 Calorimeter Group, B.~Andrieu et al.,
Nucl. Instr. and Meth. A350 (1994) 57;
{\it idem}, Nucl. Instr. and Meth. A336 (1993) 499.
\vspace{-2mm}
 
 
\bibitem{LEGOSUSS}
LEGO~0.02 and SUSSEX~1.5;
K.~Rosenbauer, Doktor Dissertation, RWTH Aachen (in German),
PITHA 95/16 (July 1995).
\vspace{-2mm}
 
\bibitem{PYTHIA}
PYTHIA~5.6;
T.~Sj\"ostrand, Comp. Phys. Comm. 39 (1986) 347; \\
T.~Sj\"ostrand and M.~Bengtsson, Comp. Phys. Comm. 43 (1987) 367.
\vspace{-2mm}
 
\bibitem{JETSET73}
JETSET~7.3;
T.~Sj\"ostrand, CERN preprint TH-6488-92 (May 1992) 284pp.
\vspace{-2mm}
 
\bibitem{MRSDM}
A.D.~Martin, W.J.~Stirling and R.G.~Roberts, Phys. Rev. D47 (1993) 867.
\vspace{-2mm}
 
\bibitem{MRSHSF}
A.D.~Martin, R.G.~Roberts and W.J.~Stirling,
Durham Univ. preprint DTP-93-86 and Rutherford Appleton Lab. preprint
RAL-93-077 (October 1993) 16pp.; \\
(PDFLIB~\cite{PDFLIB}
 nucleon structure function type 1, group 3,
 set 36) .
\vspace{-2mm}
 
\bibitem{PDFLIB}
H.~Plothom-Besch, CERN-PPE Parton Density Functions program W5051.
\vspace{-2mm}
 
\bibitem{BARTL}
A.~Bartl, H.~Fraas and W.~Majerotto, Z. Phys. C30 (1986) 441.
\vspace{-2mm}
 
\bibitem{DJANGO}
DJANGO~2.1;
G.A.~Schuler and H.~Spiesberger,
Proceedings of the Workshop Physics at HERA,
W.~Buchm\"uller and G.~Ingelman (Editors),
(October 1991, DESY-Hamburg), vol. 3 p. 1419.
\vspace{-2mm}
 
\bibitem{INGELMAN}
LEPTO~6.1;
G.~Ingelman,
Proceedings of the Workshop Physics at HERA,
W.~Buchm\"uller and G.~Ingelman (Editors),
(October 1991, DESY-Hamburg), vol. 3 p. 1366.
\vspace{-2mm}
 
\bibitem{HERACLES}
HERACLES 4.4;
A.~Kwiatkowski, H.~Spiesberger and H.-J.~M\"ohring,
Comput.~Phys.~Commun. 69 (1992) 155.
\vspace{-2mm}
 
\bibitem{ARIADNE}
ARIADNE 4.0;
L.~L\"onnblad, Comput.~Phys.~Commun. 71 (1992) 15.
\vspace{-2mm}
 
\bibitem{JETSET74}
JETSET~7.4;
T.~Sj\"ostrand, Lund Univ. preprint LU-TP-95-20 (August 1995) 321pp;
{\it idem}, CERN preprint TH-7112-93 (February 1994) 305pp.
\vspace{-2mm}
 
\bibitem{HERASF}
H1 Collaboration, I.~Abt et al., Nucl. Phys. B407 (1993) 515, \\
ZEUS Collaboration, M.~Derrick et al., Phys. Lett. B316 (1993) 412.
\vspace{-2mm}
 
\bibitem{SFGRVGLO}
M.~Gl\"uck, E.~Reya and A.~Vogt, Phys. Rev. D45 (1992) 3986;
{\it idem}, Phys. Rev. D46 (1992) 1973; 
(PDFLIB~\cite{PDFLIB}
nucleon structure function type 1, group 5 set 4;
photon structure function type 3 group 5 set 3).
\vspace{-2mm}
 
\bibitem{H1CALEPI}
H1 Calorimeter Group, B. Andrieu et al.,
Nucl. Inst. and Meth. A344 (1994) 492.
\vspace{-2mm}
 
\bibitem{MUEVENT}
H1 Collaboration, T.~Ahmed et al.,
``Observation of an $e^+ p \rightarrow \mu^+ X$
  Event with High Transverse Momenta at HERA'',
DESY preprint 94-248 (December 1994) 9pp.
\vspace{-2mm}
 
\bibitem{BAUR}
U.~Baur, J.A.M.~Vermaseren, D.~Zeppenfeld, Nucl. Phys. B375 (1992) 3.
\vspace{-2mm}
 
\bibitem{KONKOKI}
In the course of completing this paper, we became aware of an
alternative \Rp\ SUSY interpretation of the observed
$ e^+ p \rightarrow \mu^+ X $ event candidate proposed by \\ 
  T.~Kon, T.~Kobayashi and S.~Kitamurama,
  Seikei Univ. preprint ITP-SU-96/02 (Tokyo, January 1996) 8pp. \\
There the event is discussed in terms of the single production
of a scalar top squark through a $\lambda'_{131}$ coupling.
The process
$ e^+ p \rightarrow b \chi_1^+ \rightarrow b \mu^+ \nu_{\mu} \chi_1^0$
with a stable $\chi_1^0$ would lead to a signature similar to that of
our S6 channel.
The process
$ e^+ p \rightarrow b \chi_1^+ \rightarrow b \mu^+ \nu_{\mu} \chi_1^0
        \rightarrow b \mu^+ \nu_{\mu} b d \nu $
would lead to a signature similar to that of our S8 channel.
\vspace{-2mm}
 
 
\bibitem{ROY}
D.P.~Roy, Phys. Lett. B283 (1992) 270.
\vspace{-2mm}
 
\bibitem{BARGER}
V.~Barger, G.~F.~Giudice and T.~Han, Phys. Rev. D40 (1989) 2987.
\vspace{-2mm}
 
\bibitem{DREINERM}
H.~Dreiner and P.~Morawitz, Nucl. Phys. B428 (1994) 31.
\vspace{-2mm}
 
\bibitem{ALEPHRP}
ALEPH Collaboration, D.~Buskulic et al., Phys. Lett. B349 (1995) 238.
\vspace{-2mm}
 
\bibitem{DREES}
M.~Drees and K.~Grassie  Z. Phys. C28 (1985) 451.
\vspace{-2mm}
 
\bibitem{HIRSCH}
M.~Hirsch, H.V.~Klapdor-Kleingrothaus and S.G.~Kovalenko,
Phys. Rev. Lett. 75 (1995) 17.
\vspace{-2mm}

\bibitem{KON}
T.~Kon, T.~Kobayashi and S.~Kitamura, Phys. Lett. B333 (1994) 263.
\vspace{-2mm}
 
\bibitem{OPAL}
OPAL Collaboration, P.~D.~Acton et al., Phys. Lett. B337 (1994) 207.
\vspace{-2mm}
%
\end{thebibliography}
\end{document}